\documentclass[conference]{IEEEtran}

\usepackage[T1]{fontenc}
\usepackage{cite}
\usepackage{amsmath,amssymb,amsfonts}
\usepackage{algorithmic}
\usepackage{graphicx, multirow}
\usepackage{textcomp}
\usepackage[dvipsnames]{xcolor}
\usepackage[hyphens]{url}
\definecolor{myblue}{RGB}{72, 98, 168}
\usepackage[colorlinks=true, allcolors=myblue]{hyperref}
\usepackage{graphicx}
\usepackage{physics}
\usepackage{soul}
\usepackage{float}
\usepackage{siunitx}
\usepackage{ulem}
\usepackage{makecell}

\newcommand{\added}[1]{#1}
\newcommand{\removed}[1]{}

\newcommand{\snaq}[0]{SNAQ}
\newcommand{\twobyN}[0]{Bilinear}
\newcommand{\spinbus}[0]{UnitCell}

\DeclareUnicodeCharacter{2009}{\,} 

\def\BibTeX{{\rm B\kern-.05em{\sc i\kern-.025em b}\kern-.08em
    T\kern-.1667em\lower.7ex\hbox{E}\kern-.125emX}}
\begin{document}

\pdfpagewidth=8.5in
\pdfpageheight=11in

\newcommand{\iscasubmissionnumber}{2698}

\pagenumbering{arabic}

\title{A manufacturable surface code architecture for spin qubits with fast transversal logic} 
\author{Jason D. Chadwick, Willers Yang, Joshua Viszlai, and Frederic T. Chong
\\University of Chicago, Chicago, IL, USA
\\\{jchadwick, willers, jviszlai, ftchong\}@uchicago.edu}

\maketitle
\thispagestyle{plain}
\pagestyle{plain}

\begin{abstract}
    Spin qubits in silicon quantum dot arrays are a promising quantum computation platform for long-term scalability due to their small qubit footprint and compatibility with advanced semiconductor manufacturing. However, spin qubit devices face a key architectural bottleneck: the large physical footprint of readout components relative to qubits prevents a dense layout where all qubits can be measured simultaneously, complicating the implementation of quantum error correction. This challenge is offset by the platform's unique rapid shuttling capability\added{, which can be used to transport qubits to distant readout ports}. In this work, we explore the design constraints and capabilities of spin qubits in silicon and propose the SNAQ (\ul{S}huttling-capable \ul{N}arrow \ul{A}rray of spin \ul{Q}ubits) surface code architecture, which relaxes the 1:1 readout-to-qubit assumption by leveraging spin shuttling to time-multiplex ancilla qubit initialization and readout. Our analysis shows that, given sufficiently high (experimentally demonstrated) qubit coherence times, \removed{this tradeoff}\added{\snaq{}} delivers an orders-of-magnitude reduction in chip area per logical qubit. Additionally, \added{by} using a denser grid of physical qubits, SNAQ enables fast transversal logic for short-distance logical operations, achieving \removed{2.0-14.6}\added{over 10}$\times$ improvement in local logical clock speed while still supporting global operations via lattice surgery. \added{This translates to a 3.2-5.7$\times$ improvement in key Adder and Lookup fault-tolerant subroutines.}\removed{This translates to a 60-64\% reduction in spacetime cost of 15-to-1 magic state distillation, a key fault-tolerant subroutine.} Our work pinpoints critical hardware metrics and provides a compelling path toward high-performance fault-tolerant computation on near-term-manufacturable spin qubit arrays.
\end{abstract}

\section{Introduction}

\begin{figure}
    \centering
    \includegraphics[width=\linewidth]{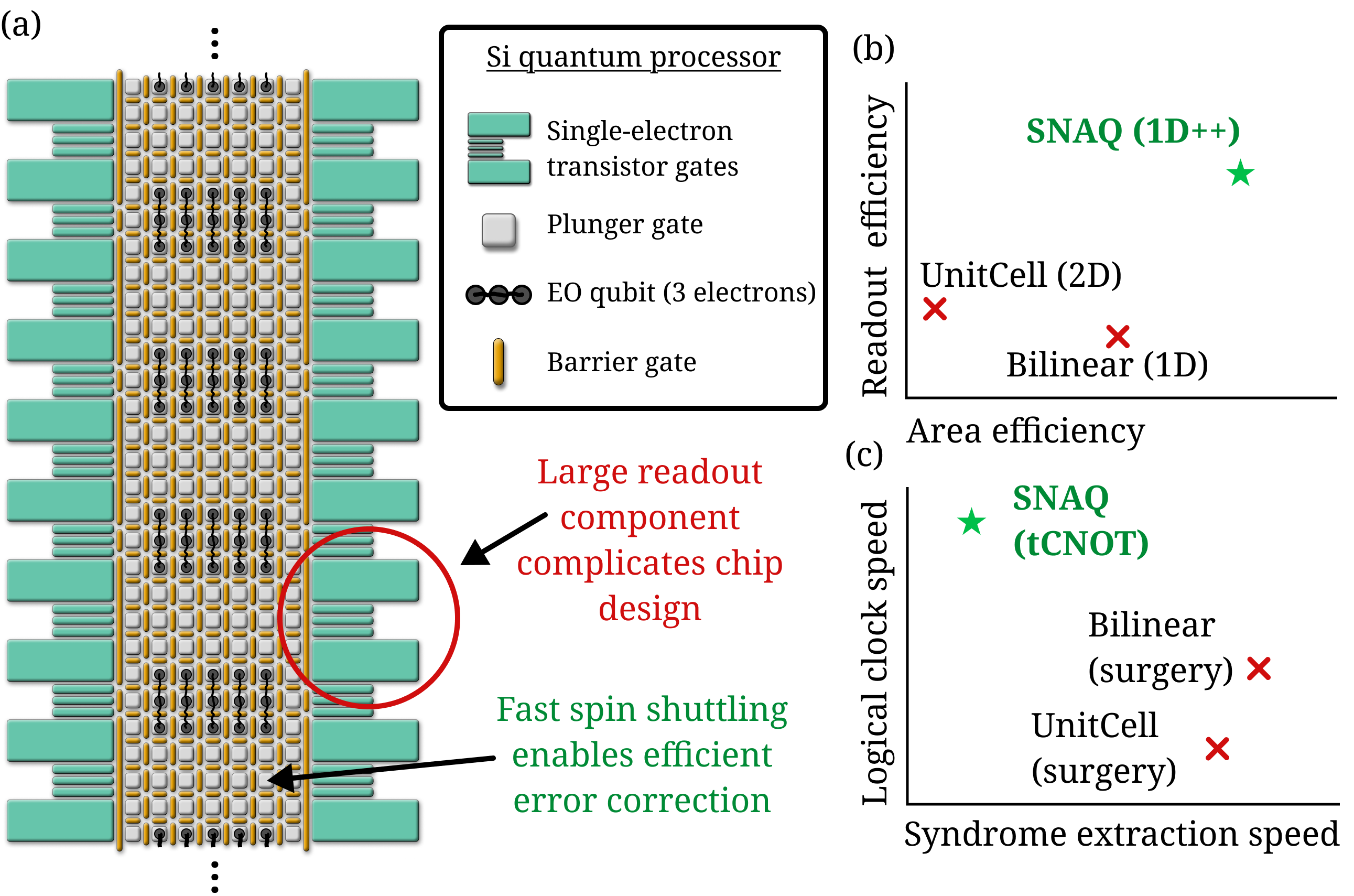}
    \caption{(a) Depiction of a possible implementation of a 7-dot-wide array of quantum dots, similar to devices fabricated by \added{industrial and academic groups} \cite{philips_universal_2022, borsoi_shared_2024, henry_2d_2025, ha_twodimensional_2025, george_12spinqubit_2025}. Electrons (black) are held in place at the plunger gates (light gray) and can interact with their neighbors or shuttle to adjacent plungers by manipulating the voltages on the barrier gates (orange). Single-electron transistors (green) allow for qubit readout on the two sides of the array. (b) \snaq{} achieves similar performance to baselines while improving chip area efficiency and \added{readout component}\removed{wiring} efficiency \added{per logical qubit}. (c) \snaq{} avoids the downside of a longer syndrome extraction cycle by enabling transversal CNOTs (tCNOTs), achieving significantly faster logical clock speeds compared to lattice-surgery-based approaches.}
    \label{fig:hero}
\end{figure}

Scaling quantum computers up to the million-qubit processors \added{needed} to enable powerful applications \cite{kim_faulttolerant_2022, beverland_assessing_2022, leblond_realistic_2024, gidney_how_2025, zhou_resource_2025} is an immense architectural challenge.\removed{and it is not yet known which hardware modality (transmons, neutral atoms, trapped ions, etc.) will ultimately be able.} Spin qubits in silicon quantum dot arrays are a promising candidate for large-scale quantum computing, primarily due to their small footprint (on the order of $100 \times 100$~\si{\nano\meter} per qubit) and their compatibility with existing advanced semiconductor fabrication techniques \cite{maurand_cmos_2016, veldhorst_silicon_2017, li_crossbar_2018, xue_cmosbased_2021, neyens_probing_2024, steinacker_300_2024}. However, the \added{small footprint} that makes spin qubits attractive creates a fundamental architectural challenge: \removed{The}physical \removed{footprint of}readout components, such as a single-electron transistor (SET) or single-electron box (SEB), require\removed{s} large reservoirs and ohmic contacts that are many times the size of a quantum dot \cite{morello_singleshot_2010, curry_singleshot_2019, connors_rapid_2020, hendrickx_singlehole_2020, oakes_fast_2023, schmidt_compact_2024}. This makes it difficult to build a dense array of dots while dedicating a unique readout component to each qubit, instead incentivizing asymmetric fixed-width arrays as shown in Figure~\ref{fig:hero}(a). On the other hand, spin qubits also offer the unique and compelling capability of extremely fast spin shuttling between quantum dots \cite{seidler_conveyormode_2022, smet_highfidelity_2024}, which can enable novel shuttling-based hardware layouts that can overcome the readout bottleneck \cite{boter_spiderweb_2022, cai_looped_2023, strikis_quantum_2023, kunne_spinbus_2024, siegel_early_2024, siegel_snakes_2025}. \added{Shuttling comes with a cost, though.}\removed{However, shuttling is not free.} Qubits accumulate errors proportional to the distance shuttled, making it crucial to minimize shuttling distance when possible.

Previous spin qubit architecture proposals have addressed these challenges by \added{introducing} sparser arrays, creating large interior spaces to fit the needed components, but at the cost of a large shuttling overhead required to enable the dense 2D qubit connectivity needed for error correction. In this work, instead of designing an entirely new hardware layout, we \added{consider the proven design of a} fixed-width\added{, limited-readout} array\removed{ as the fundamental building block} \cite{philips_universal_2022, borsoi_shared_2024, henry_2d_2025, ha_twodimensional_2025}, posing the central question: can we efficiently run the surface code in this constrained geometry \removed{by trading parallel measurement}\added{by sacrificing measurement parallelism in exchange} for a denser qubit array?

We propose a hardware-aware surface code architecture named \snaq{} (\ul{S}huttling-capable \ul{N}arrow \ul{A}rray of spin \ul{Q}ubits) that leverages fast spin shuttling to overcome the limited readout capabilities in silicon quantum dot hardware. We show that for sufficiently low idle error rates demonstrated by existing spin qubit hardware, the initialization and measurement of physical qubits can be serialized with manageable impact on the logical performance of the surface code. This creates a significantly denser qubit array, which reduces the required amount of spin shuttling \added{for error correction} and enables the transversal CNOT for short-range logical multiqubit gates, delivering logical clock speed improvements and \removed{potentially improved}\added{greater potential} parallelism. This work makes the following main contributions:

\begin{figure}
    \centering
    \includegraphics[width=\linewidth]{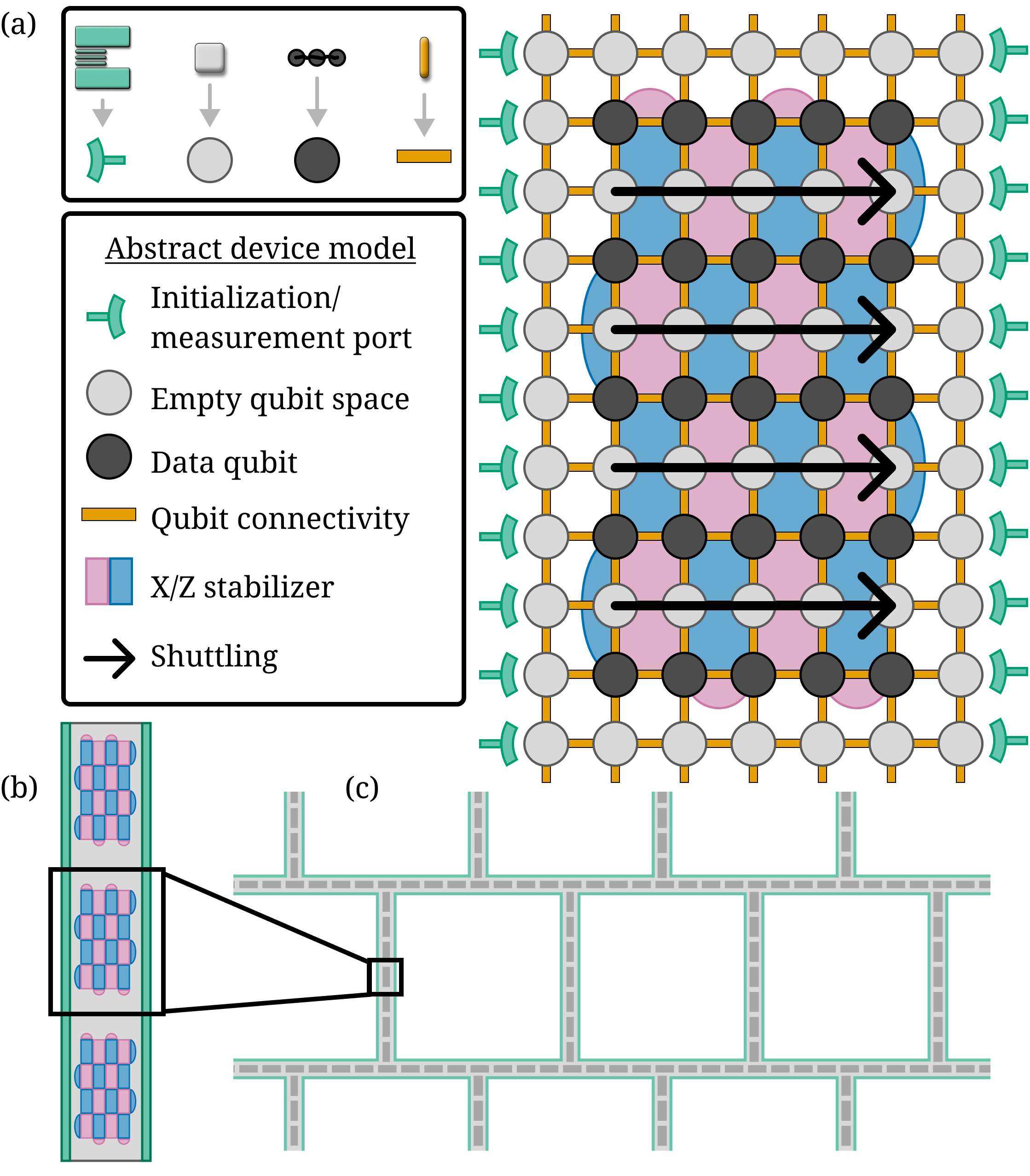}
    \caption{Proposed \snaq{} (\ul{S}huttling-capable \ul{N}arrow \ul{A}rray of spin \ul{Q}ubits) architecture. (a) For ease of visualization, we translate from the specific chip components in Figure~\ref{fig:hero} to the abstract components we will use in the rest of the paper. Depicted is a layout of a distance-5 surface code patch on a 7-dot-wide array. Initialization and readout components (green) are only available on the edges of the array. Data qubits (dark gray) are interleaved with channels of empty dots (light gray) used to shuttle surface code ancilla qubits from edge to edge in the direction of the black arrows. Surface code X and Z stabilizers (blue and pink) are shown behind the array, supported on the data qubits. (b) A logical 1 $\times$ N layout of surface codes in a narrow array. (c) Possible design of a fully scalable architecture consisting of loops of \snaq{} channels, with large spaces in between to allow for integration of control wires and readout.}
    \label{fig:snaq}
\end{figure}

\begin{itemize}
    \item We identify and characterize the qubit-to-readout area mismatch as a key design challenge for scalable spin qubit architectures.
    \item We propose SNAQ, a novel surface code architecture that leverages fast spin shuttling to time-multiplex syndrome extraction and overcome this bottleneck in near-term-manufacturable arrays.
    \item We analyze the physical parameters for this new design with detailed circuit-level simulations, pinpointing the readout density $\rho$ and the qubit idle error rate $p_\text{id}$ as the most critical hardware metrics for its viability. Although the idle error restricts the lowest achievable logical error rate\added{s}, our method shows competitive performance against baselines for algorithmically relevant error rates.
    \item We demonstrate through simulation that \snaq{} enables low-latency transversal CNOTs (tCNOTs) for local multiqubit gates, creating a two-tiered latency hierarchy between fast-local tCNOTs and slow-global lattice surgery.
    \item \added{We perform a thorough performance evaluation, showing that \snaq{} can achieve over 10$\times$ faster logical clock speed. This translates to 57-84\% reduced cost for key fault-tolerant primitives, enabling an estimated 2.4-4.2$\times$ improvement in resource estimates for large-scale factoring algorithms.}\removed{We quantify the benefits of the fast-local latency tier, demonstrating a \removed{2.0-14.6}\added{4.0-22.3}$\times$ speedup per operation and a \removed{60-64}\added{57-84}\% spacetime cost reduction for \added{key fault-tolerant primitives, which translate to 2.4-4.2$\times$ improvements in resource estimates for large-scale factoring algorithms.}\removed{ the 15-to-1 magic state distillation benchmark.}}
\end{itemize}
  
\section{Background}\label{sec:background}

In this section, we explain the components and terminology of silicon quantum dot devices and provide some high-level background on quantum error correction and the surface code. We \removed{urge}\added{encourage} the reader to refer to Ref.~\cite{burkard_semiconductor_2023} for a more thorough background on spin qubits and to Refs.~\cite{dennis_topological_2002, fowler_surface_2012} for more information on the surface code.

\subsection{Spin qubits in silicon}

Semiconductor spin qubits generally refer to nanoscale qubits encoded in the spin states of electrons or holes in a semiconductor substrate. In this work, we will specifically consider electrons held in quantum dots in silicon, which is a platform that has seen considerable recent progress \cite{burkard_semiconductor_2023, weinstein_universal_2023, george_12spinqubit_2025, wu_simultaneous_2025, broz_demonstration_2025, steinacker_industrycompatible_2025}. Here we will describe the general terminology and architectural implications of the silicon spin qubit platform, leaving a specific discussion of state-of-the-art performance metrics to Section~\ref{sec:hw-req}. From an architectural perspective, these devices present a compelling substrate for large-scale fault-tolerant quantum computing: the qubit footprint is comparably small (dot pitches on the order of tens of nanometers), fabrication is compatible with \added{existing} CMOS \removed{semiconductor}processes, and spin shuttling is a novel and powerful tool. \added{We refer the interested reader to Ref. \cite{burkard_semiconductor_2023} for a more thorough review of semiconductor spin qubits.}

A silicon quantum dot array is defined by electrostatic gates as depicted in Figure~\ref{fig:hero}, in \added{which} we identify three important categories of these gates: ``plunger'' gates, each of which is tuned to create a quantum dot that can hold one electron; ``barrier'' gates, which control inter-dot tunnel coupling, used for shuttling and for electron-electron \added{entanglement}\removed{interaction} via the exchange interaction; and the gates \added{that define the large electron reservoirs and quantum dot that form a single-electron transistor, which is used to measure the spin states of electrons}\removed{temp}. The simplest way to \added{compute with electrons in quantum dots}\removed{use a quantum dot array as a quantum processor} is to treat individual electrons as qubits, where the two logical states are the spin states $\ket \uparrow$ and $\ket \downarrow$; alternative encodings can exploit up to four electrons per qubit to trade control complexity, magnetic gradient requirements, and coupling speed \cite{burkard_semiconductor_2023}. Generally, increasing the number of electrons that encode each qubit simplifies the control requirements\added{, but may reduce best achievable qubit coherence}. The exchange-only (EO) qubit encoding uses three electrons to define each qubit, and is a promising choice for scalability because it relaxes the requirement for localized magnetic field control and instead can be operated by only tuning DC voltages on barriers and plungers \cite{bacon_universal_2000, divincenzo_universal_2000}. In all qubit encodings, logical operations between qubits are mediated by the exchange interaction: by tuning the inter-dot tunnel barrier and dot detuning, one can dynamically adjust the exchange coupling \added{between two adjacent electrons} and thereby implement controlled-phase, CNOT-equivalent, or SWAP-like operations. It may even be beneficial to use multiple different encodings in one architecture \cite{gutierrez_comparison_2025}.

\added{Silicon quantum processors introduce several unique architectural considerations that we account for in this work.}\removed{The architectural implications of silicon quantum processors diverge from more familiar transmon, neutral atom, or trapped-ion technologies in several key ways.} First, readout of spin qubits typically uses spin-to-charge conversion, which \added{is implemented using components that are significantly larger than those that hold the qubits}\removed{tends to require a significantly larger component area than the qubit} \cite{morello_singleshot_2010, connors_rapid_2020, hendrickx_singlehole_2020, curry_singleshot_2019, oakes_fast_2023, schmidt_compact_2024}. Designing the layout of a silicon spin processor is thus similar to the ``pitch matching'' exercise studied extensively in classical architecture, where physical components of different sizes must be \added{integrated into a cohesive} layout \cite{alma998987573606533, oskin_building_2003, weste_cmos_2011, seyyedaghaei_memorycentric_2025, iyer_multitier_2025}. Because the qubit-qubit interactions are fundamentally constrained to be short-distance, the large readout size \added{has motivated the current approach of placing readout sensors on the edges of a quantum dot array to allow for a dense qubit grid, which has enabled impressive experimental demonstrations \cite{philips_universal_2022, borsoi_shared_2024, henry_2d_2025, ha_twodimensional_2025} but has not previously been expected to scale to the fault-tolerant era.}\removed{often results in architectures in which readout sensors are sparsely distributed or placed on the edges, and  the internal qubit\added{s} must rely on qubit shuttling for readout access.} Second, the capability of ``shuttling'' (moving an electron spin coherently between quantum dots) has a different cost model than qubit movement in neutral atoms or trapped ions. In spin qubits, the shuttling is very fast (a few nanoseconds per dot-to-dot transfer), but shuttling errors accumulate more rapidly with distance than in atomic systems.

Silicon spin qubit technology is a highly promising candidate for truly scalable quantum processors. At the same time, its distinct challenges and opportunities call for novel architectural ideas to make the best use of this hardware and enable efficient fault-tolerant computation.

\subsection{Quantum error correction and the surface code}

Quantum error correction (QEC) is the process by which many physical qubits are used to encode a smaller number of logical qubits with much higher lifetimes and operating fidelities. A typical QEC code is encoded into a large number of \textit{data qubits} and operates by repeatedly measuring a set of stabilizers of the code, which are joint observables on multiple data qubits. Typically, each stabilizer is measured using an \textit{ancilla qubit} that interacts with each of the relevant data qubits before being measured. Measurement of all stabilizers of the code produces a\added{n error} syndrome that indicates the difference between the expected stabilizer parities and the observed values. This syndrome must then be decoded by a classical algorithm to determine which physical qubits experienced errors. We will refer to this stabilizer measurement process as syndrome extraction (SE). Repeatedly performing SE rounds and decoding allows individual physical qubit errors to be corrected, stabilizing the logical qubit and extending its lifetime.

Important metrics of a QEC code in the context of this work are the code distance $d$, which determines the error robustness and size of a code, and the threshold $p^*$, which is the physical error rate below which the code can be scaled up to increase logical lifetimes. When the physical error rate $p$ is below $p^*$, the logical error rate $p_L$ scales approximately as
\begin{equation}
    p_L \approx A\bigg(\frac{p}{p^*}\bigg)^{(d+1)/2}.
    \label{eq:pL}
\end{equation}

Among the many QEC codes proposed, the rotated surface code stands out for its easy-to-build planar connectivity requirements, relatively high threshold, ease of decoding \cite{dennis_topological_2002, fowler_surface_2012}, and well-understood logical operations \cite{horsman_surface_2012, fowler_low_2019, litinski_game_2019}. One logical qubit in a distance-$d$ surface code is encoded on a square $d \times d$ grid of physical qubits\added{, which we will refer to as a ``surface code patch''}. Each stabilizer within this grid is either a joint X or Z operator supported on four neighboring qubits (or two on a boundary). Including edge stabilizers, there are $d^2-1$ stabilizers that must be measured in each SE round. \added{Conventionally, each of these stabilizers is measured using a unique physical ancilla qubit which is initialized, entangled with the involved data qubits, and measured. Serialization of syndrome extraction has been studied before to reduce physical qubit overheads \cite{sato_scheduling_2025, ye_quantum_2025} or to adapt to control constraints \cite{helsen_quantum_2018}; in this work, we study the forced serialization of the measurement of ancilla qubits due to the limited readout capacity of the chip.}\removed{Multiqubit logical operations in the surface code can be performed in two primary ways, through lattice surgery or transversal CNOTs, both of which are supported on \snaq{}.}

\added{Two main methods have been proposed to entangle logical surface code qubits, both of which are supported in \snaq{}. Hardware that is restricted to nearest-neighbor connectivity can perform lattice surgery \cite{horsman_surface_2012, fowler_low_2019, litinski_game_2019}, which involves the ``merging'' and ``splitting'' of surface code patches to perform multiqubit Pauli measurements, each of which takes $d$ SE rounds. Alternatively, hardware with long-range connections can perform a transversal CNOT (tCNOT), where each data qubit of one patch interacts with its corresponding data qubit in the second patch. When combined with specialized decoding techniques \cite{cain_correlated_2024, zhou_algorithmic_2024}, tCNOTs only require $\Theta(1)$ SE rounds after each operation. Prior work has found that less than one SE round is needed per tCNOT \cite{cain_correlated_2024}, saving significant temporal costs compared to lattice surgery. Additionally, while parallel shuttling of qubits is possible in a transversal computation, parallel lattice surgery through the same shared routing space is not, so tCNOTs may have compilation benefits for certain workloads in restricted logical layouts. Decoding a tCNOT-based logical circuit ostensibly requires use of a more complex decoder than in a lattice-surgery-based system, but methods have recently been developed to reduce the transversal decoding problem to one closely resembling an idling logical qubit, allowing for fast decoding \cite{cain_fast_2025}.}

\section{Hardware requirements}\label{sec:hw-req}

\snaq{} is designed to be compatible with current or near-future silicon spin qubit technology. This section reviews state-of-the-art spin qubit capabilities and lists the concrete device-level capabilities we assume for \snaq{} and the experimental evidence that these capabilities are achievable \added{in the near term.}\removed{with current or near-future silicon spin qubit technology.}

\subsection{Array geometry and readout placement}
A \snaq{} device consists of a narrow and dense two-dimensional grid of quantum dots with readout components on two edges of the array, such as the 7-dot-wide example shown in Figure~\ref{fig:hero}(a). This design maximizes compatibility with existing semiconductor fabrication techniques \cite{elsayed_low_2024, george_12spinqubit_2025}, and is a directly scaled-up version of existing devices such as Intel's 4x27-dot array \cite{henry_2d_2025} or HRL's 3x3-dot array \cite{ha_twodimensional_2025}. The choice of a fixed width means that the array can be constructed using a fixed number of interconnect layers, which are required to control the interior dots in the array, regardless of the size along the second axis of the array. A key parameter that affects the performance of \snaq{} is the \textit{readout density}, $\rho$, which is the density of readout devices along the sides of the array relative to the rows of qubits.\removed{We will discuss this parameter in more detail in Section~\ref{sec:snaq-layout}.} Readout density can be increased by building more complex routing on the array edges or by using a qubit encoding that occupies more dots, such as the three-electron exchange-only qubits depicted in Figure~\ref{fig:hero}.

\subsection{Spin shuttling}
\snaq{} relies on coherent electron transport to move ancilla qubits between readout/initialization regions and interior interaction sites. There are two distinct methods for shuttling an electron between dots. ``Bucket-brigade'' shuttling is performed by a series of discrete single-dot jumps, which has the advantage of only requiring discrete control pulses on plungers and barriers, and has achieved error rates of less than 0.3\% per hop \cite{zwerver_shuttling_2023, noiri_shuttlingbased_2022}. ``Conveyor-mode'' shuttling involves operating the metal gates along the path with smoothly oscillating voltages to engineer a continuously moving potential wave that carries the electron along\removed{ the path}. This second approach is more complex to engineer, but can reach faster shuttling rates and has been achieved with per-hop fidelities of $99.99\%$\removed{ in experiment} \cite{smet_highfidelity_2024}. At a nominal dot pitch of 100~\si{\nano\meter}, these results correspond to per-dot shuttling latencies around 2~\si{\nano\second}. We find that these experimentally achieved speeds and fidelities are already sufficient to enable effective quantum error correction, and we expect further improvement in the future.

\subsection{Coherence}
Data qubits in \snaq{} are idle for longer intervals than a standard surface code due to serialized ancilla initialization and measurement. The idling coherence time of the data qubits is therefore a crucial parameter that will significantly affect the performance of a \snaq{} device. \removed{Quantum-dot electron-spin}\added{Achieved spin qubit} coherence times vary significantly between device types and choice of qubit encoding. State-of-the-art coherence times range from around 1~\si{\milli\second} to 0.56~\si{\second} for one and two-electron qubits \cite{muhonen_storing_2014, veldhorst_addressable_2014, malinowski_notch_2017, laucht_dressed_2017}. Three-electron EO qubits, whose appeal has grown recently, have been stabilized along one axis for up to 720~\si{\micro\second} \cite{sun_fullpermutation_2024}. Coherence is expected to improve further as device uniformity and material purity improve \cite{burkard_semiconductor_2023}. We find that a coherence time of at least 200~\si{\micro\second} is required to enable competitive error correction on \snaq{}, with the specific number depending on the code distance, readout element density, and the strength of other error sources.

\subsection{Physical qubit gate fidelities}
\snaq{} does not impose any unique constraints on the fidelity of single- and two-qubit gates compared to surface code proposals on other hardware modalities, simply requiring that the gate fidelity is below the threshold of the surface code (typically around $10^{-2}$). We set a realistic gate error target of $10^{-3}$, which has been nearly reached or \added{surpassed} in many experimental spin qubit demonstrations \cite{takeda_resonantly_2020, mills_twoqubit_2022, weinstein_universal_2023, wu_simultaneous_2025, broz_demonstration_2025, steinacker_industrycompatible_2025}.

\subsection{Initialization and readout}

Spin qubit readout via spin-dependent tunneling \cite{elzerman_singleshot_2004, morello_singleshot_2010} offers a promising path towards fast high-fidelity readout. Early experiments have demonstrated readout fidelity above 99\% in 6~\si{\micro\second} with a single-electron box (SEB) \cite{oakes_fast_2023}, 1.6~\si{\micro\second} using a single-electron transistor (SET) \cite{connors_rapid_2020}, and 990~\si{\nano\second} using a dot-charge sensor \cite{blumoff_fast_2022}. Theoretical analyses predict that a readout fidelity of 99.97\% and duration as short as 100~\si{\nano\second} could be achieved with the SEB \cite{oakes_fast_2023}, and readout above 99\% fidelity could be performed in only 36~\si{\nano\second} with fully optimized SET device parameters \cite{keith_singleshot_2019}. In this work, we assume that initialization and readout each take 500~\si{\nano\second} and have fidelities comparable to physical quantum gates.

\subsection{Practical summarized requirements for near-term \snaq{}}
The following target parameters capture the regime in which serialized readout with shuttling yields competitive logical performance, as we will show in simulation in the following sections:
\begin{itemize}
\item \textbf{Array geometry:} Dense fixed-width rectangular array of nearest-neighbor-connected dots with readout elements confined to edges \cite{henry_2d_2025, george_12spinqubit_2025, ha_twodimensional_2025}. Readout density $\rho$ at least 1, ideally closer to 2.
\item \textbf{Shuttling:} Per-hop error $p_\text{sh} \le 10^{-4}$ at 2~\si{\nano\second} latency \cite{smet_highfidelity_2024}.
\item \textbf{Idling:} Coherence time exceeding 200~\si{\micro\second} (effective idle-error-per-\si{\micro\second} $p_{\mathrm{id}} \lesssim 5 \times 10^{-3}$) under active stabilization such as dynamical decoupling \cite{muhonen_storing_2014, veldhorst_addressable_2014, malinowski_notch_2017, laucht_dressed_2017, sun_fullpermutation_2024}.
\item \textbf{Physical logic gates:} CNOT error $p_\text{g} \gtrsim99.9\%$ \cite{takeda_resonantly_2020, mills_twoqubit_2022, weinstein_universal_2023, wu_simultaneous_2025, broz_demonstration_2025, steinacker_industrycompatible_2025}.
\item \textbf{Initialization and measurement:} Initialization and readout fidelity equal to $p_\text{g}$ with 500~\si{\nano\second} latency each \cite{keith_singleshot_2019}.
\end{itemize}

\added{While the state-of-the-art metrics cited above have been demonstrated across different devices, they rely on compatible semiconductor fabrication techniques. We perform sensitivity studies to variations in each of these variables, finding \snaq{} to be competitive across wide parameter ranges.}\removed{To provide a clear characterization of the architecture, we restrict our evaluations to the most impactful regions of the design space, prioritizing sensitivity studies of the variables that drive the primary tradeoffs.}

\begin{figure*}[tp]
    \centering
    \includegraphics[width=\linewidth]{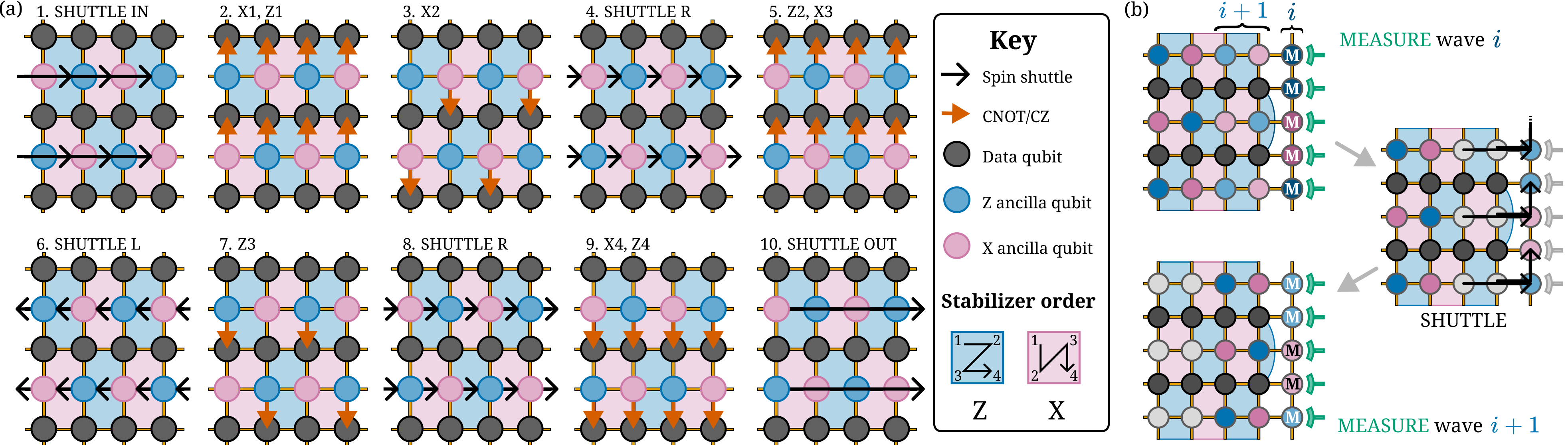}
    \caption{(a) Schedule of shuttle and CNOT/CZ operations that implement the surface code X and Z checks in \removed{the required}\added{a distance-preserving} order. Each ancilla qubit is responsible for interacting with four nearby data qubits, and it must interact in a specific order to avoid hook errors \cite{dennis_topological_2002}, as shown in the lower right. (b) Serialized measurement of a group of ancilla qubits for readout density $\rho=1$. Ancilla qubits in group $i$ are measured first, followed by qubits in group $i+1$.}
    \label{fig:cx_order}
\end{figure*}

\section{The \snaq{} architecture}\label{sec:snaq}

The layout of our proposed architecture is shown in Figure~\ref{fig:snaq}(a), where a \added{7}-qubit wide array is used to support a \added{$d=5$} surface code.

\subsection{Hardware layout}\label{sec:snaq-layout}

\snaq{} consists of a dense fixed-width array of quantum dots with readout components on both sides, as shown in Figure~\ref{fig:hero}(a). Although previous work has proposed building long shuttling channels to route around these large readout zones \cite{kunne_spinbus_2024, otxoa_spinhex_2025}, in this work we explore the more near-term friendly approach of placing readout elements along two outer edges of a fixed-width array \cite{henry_2d_2025}, keeping the interior free for uniform dot placement and nearest-neighbor exchange coupling. A fixed array width $w$ limits the largest surface code that we can embed in the array using the mapping of Figure~\ref{fig:snaq}(a) to $d_\text{max}=w-2$. A two-column layout to \added{enable} multiqubit interactions via lattice surgery would support a maximum distance of $d_\text{max} = \lceil \frac{w-3}{2} \rceil$.

\added{The denser qubit grid of \snaq{} compared to alternative architectures introduces additional control concerns, most importantly crosstalk between nearby control lines when addressing qubits in parallel. However, this is not a fundamental obstacle, as recent work has shown that, with proper calibration of crosstalk compensation terms, voltage pulses can be applied in parallel in very close range \cite{madzik_operating_2025}.}

The readout density $\rho$, the ratio of readout sensors to array rows, is a key hardware parameter. A density of 1 means that each row has a readout sensor, as \removed{in the device }shown in Figure~\ref{fig:snaq}(a). The readout density directly determines the amount of serialization required to initialize or measure some fixed number of spin qubits, so it has a significant effect on the speed and error rate of the surface code. Achieving $\rho \gg 1$ in a manufacturable device is difficult, but a small constant-factor increase (such as $\rho=2$) may be attainable with careful routing.


\subsection{Scheduling a surface code on \snaq{}}

Figure \ref{fig:cx_order}(a) shows the specific schedule of shuttles and CNOTs that perform a stabilizer measurement cycle of the surface code. The ancilla qubits are first initialized on the left side of the array and are shuttled into place as they become ready. The ancilla qubits and data qubits then perform several layers of CNOTs. Finally, the ancilla qubits are shuttled to the right side of the array, where they can be measured. Note that the depiction of shuttle operations here is slightly simplified; in reality, buffer space would be needed in between adjacent ancilla qubits for them to both be shuttled at the same time. This means that the group shuttle operation cannot be fully parallelized.

\removed{In many quantum error correction codes, including the rotated surface code, the specific schedule of CNOT gates is crucial, as some orderings of these gates can allow hook errors, where a single error on an ancilla qubit can propagate to multiple data qubit errors during a round of syndrome extraction \cite{dennis_topological_2002}. This would reduce the effective code distance, so care must be taken to choose a gate ordering that avoids hook errors. }\added{To avoid harmful ``hook errors'', where errors propagate in detrimental patterns \cite{dennis_topological_2002},}\removed{In our schedule,} we implement the \added{commonly chosen CNOT} ordering shown in the lower right of Figure~\ref{fig:cx_order}(a), where the number in each corner of a stabilizer indicates the order of the gates from the perspective of the ancilla qubit. \added{Ancilla qubits on the edges of the patch perform the same shuttling pattern, simply skipping some CNOT gates.}

\begin{figure}
    \centering
    \includegraphics[width=\linewidth]{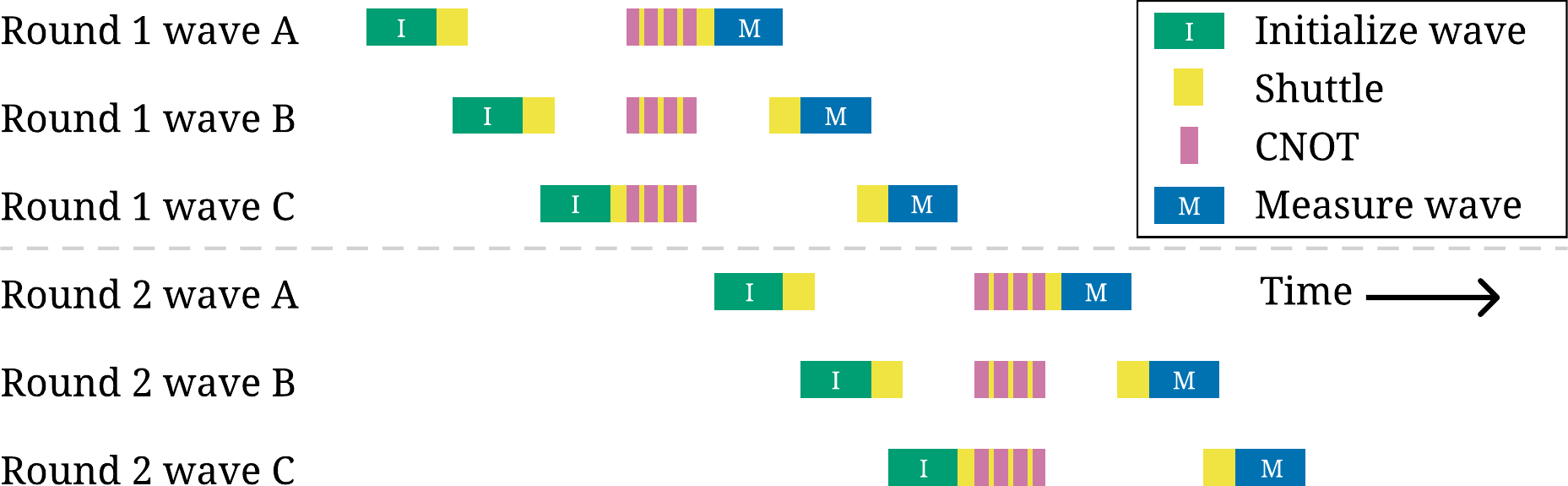}
    \caption{Example pipeline schedule showing initialization, data entanglement, and measurement for two successive rounds of ancilla qubits, split into three waves each. The measurement of the first round of ancillae can be done simultaneously with the initialization of the second round. Any blank space in the schedule indicates idling, during which dynamical decoupling techniques could be used to improve coherence.}
    \label{fig:pipelining}
\end{figure}

The primary bottleneck of the \snaq{} surface code is the initialization and measurement of $d^2-1$ ancilla qubits in each round of syndrome extraction. In the proposed syndrome extraction schedule, these ancilla qubits are all initialized on the left side of the array, shuttled into place in the interior, and then eventually measured and removed on the right side of the array. Because a given surface code patch will only have $O(d)$ available initialization/measurement devices on one of its edges for any constant density $\rho$, a sufficiently large code distance will require multiple serial waves of ancilla qubit initialization/measurement, as depicted in Figure~\ref{fig:cx_order}(b). The number of ancilla waves is described by
\begin{equation}
    n_w = \bigg\lceil \frac{d^2-1}{2 \rho (d+1)}\bigg\rceil,
    \label{eq:nw}
\end{equation}
where $d^2-1$ is the total number of ancilla qubits in a distance-$d$ surface code and $2\rho (d+1)$ is the number of readout ports on one side of a \snaq{} surface code. This serialization is the primary novel source of error that the \snaq{} surface code must mitigate. Logical initialization and measurement of a surface code require serialized readout and measurement of the $d^2$ data qubits as well, which can be done in a similar method to the ancilla qubits (and can be done in roughly half the number of waves by using both edges of the array at the same time).

Importantly, we note that steps 1 and 10 of Figure~\ref{fig:cx_order}(a), which are by far the longest due to serialization, can be pipelined: while layers of ancillae are being moved to the readout locations on the right side of the array, fresh waves of ancillae can be initialized at the same time to fill in the empty spots in the array, as shown in Figure~\ref{fig:pipelining}. Our circuit-level simulations account for this pipelining.


\subsection{Multiqubit logical gates}

\begin{figure}
    \centering
    \includegraphics[width=\linewidth]{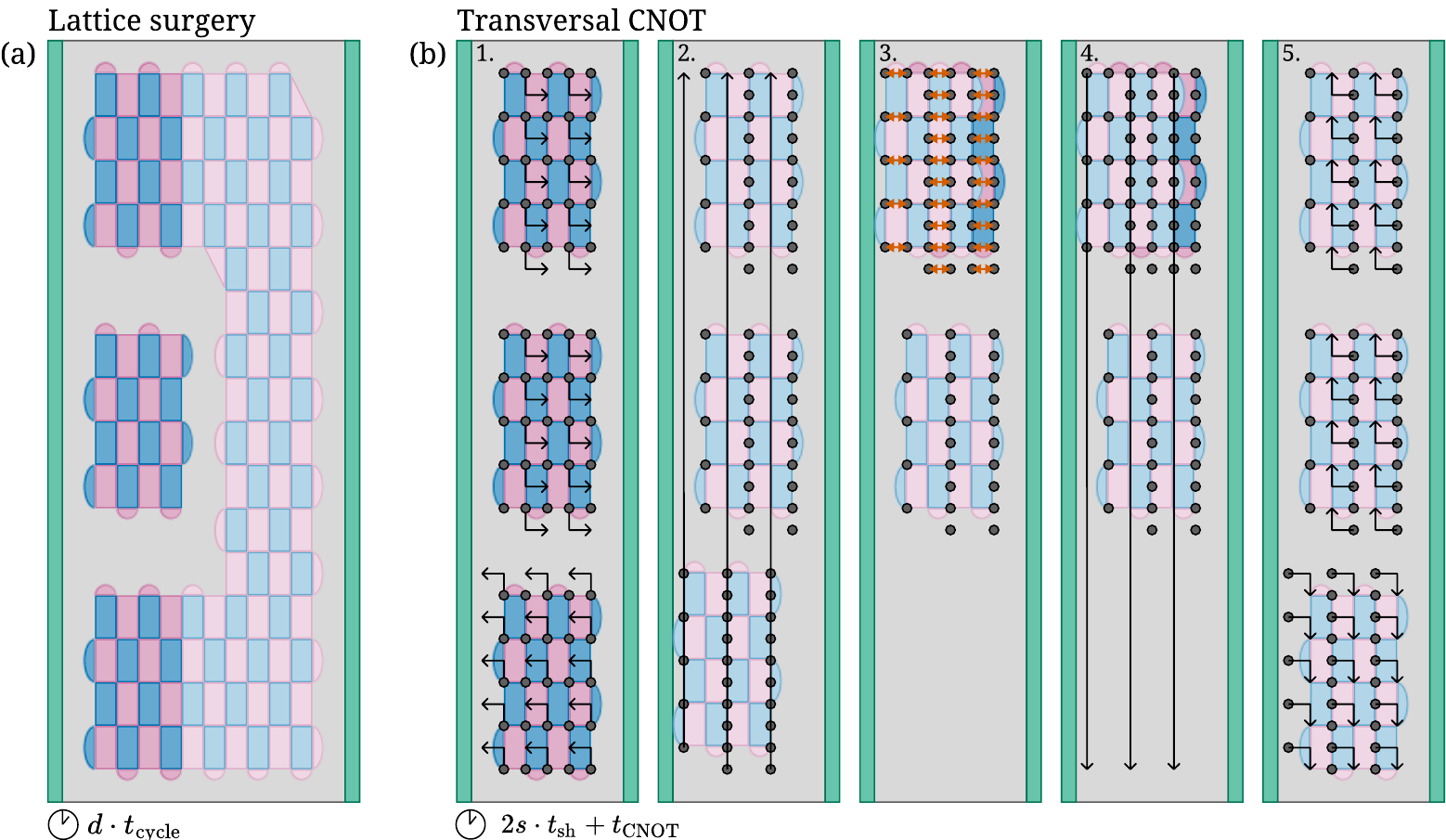}
    \caption{Two possible implementations of multiqubit operations in the \snaq{} architecture. (a) Lattice surgery can connect distant logical qubits via a channel of ``routing patches'' (lighter shaded stabilizers), which requires a wider dot array to support a second column of surface codes. (b) Spin shuttling enables transversal CNOT operations for sufficiently close tiles. Between stabilizer measurement rounds, physical qubits can shuttle past intermediate tiles to reach the target tile, then perform the transversal operation and shuttle back.}
    \label{fig:multiqubit_gates}
\end{figure}

\removed{Two main methods have been proposed to entangle logical surface code qubits. Hardware that is restricted to nearest-neighbor connectivity can perform lattice surgery \cite{horsman_surface_2012, fowler_low_2019, litinski_game_2019} (LS), which involves the merging and splitting of surface code patches to perform multiqubit Pauli measurements. In addition, hardware with long-range connections can perform a transversal CNOT (tCNOT), where each data qubit of one patch interacts with its corresponding data qubit in the second patch. Using specialized decoding techniques \cite{cain_correlated_2024, zhou_algorithmic_2024}, transversal CNOTs only require $\Theta(1)$ SE rounds after each operation, saving significant temporal costs compared to lattice surgery (the actual number can be less than 1 SE round per tCNOT \cite{cain_correlated_2024}). Additionally, while parallel shuttling of qubits is possible, parallel lattice surgery through the same shared routing space is not, so transversal CNOTs may have compilation benefits for certain workloads.} Both lattice surgery and transversal operations are supported in the \snaq{} architecture, and \added{their implementations} are depicted in Figure \ref{fig:multiqubit_gates}. Importantly, a two-column logical layout is required to support lattice surgery, while a processor that only uses tCNOTs can be operated in a single-column layout, \added{which allows twice as large a code distance on a chip of the same width.}\removed{which may be significantly easier to manufacture for the same target code distance.} The tradeoff to using tCNOTs is that they natively only work well for relatively small separation distances before shuttling errors accumulate. \removed{We explore this and suggest possible methods to extend the tCNOT's range in Section~\ref{sec:tcnot-fid}.}

\subsection{Prior proposals}

\begin{figure}
    \centering
    \includegraphics[width=0.6\linewidth]{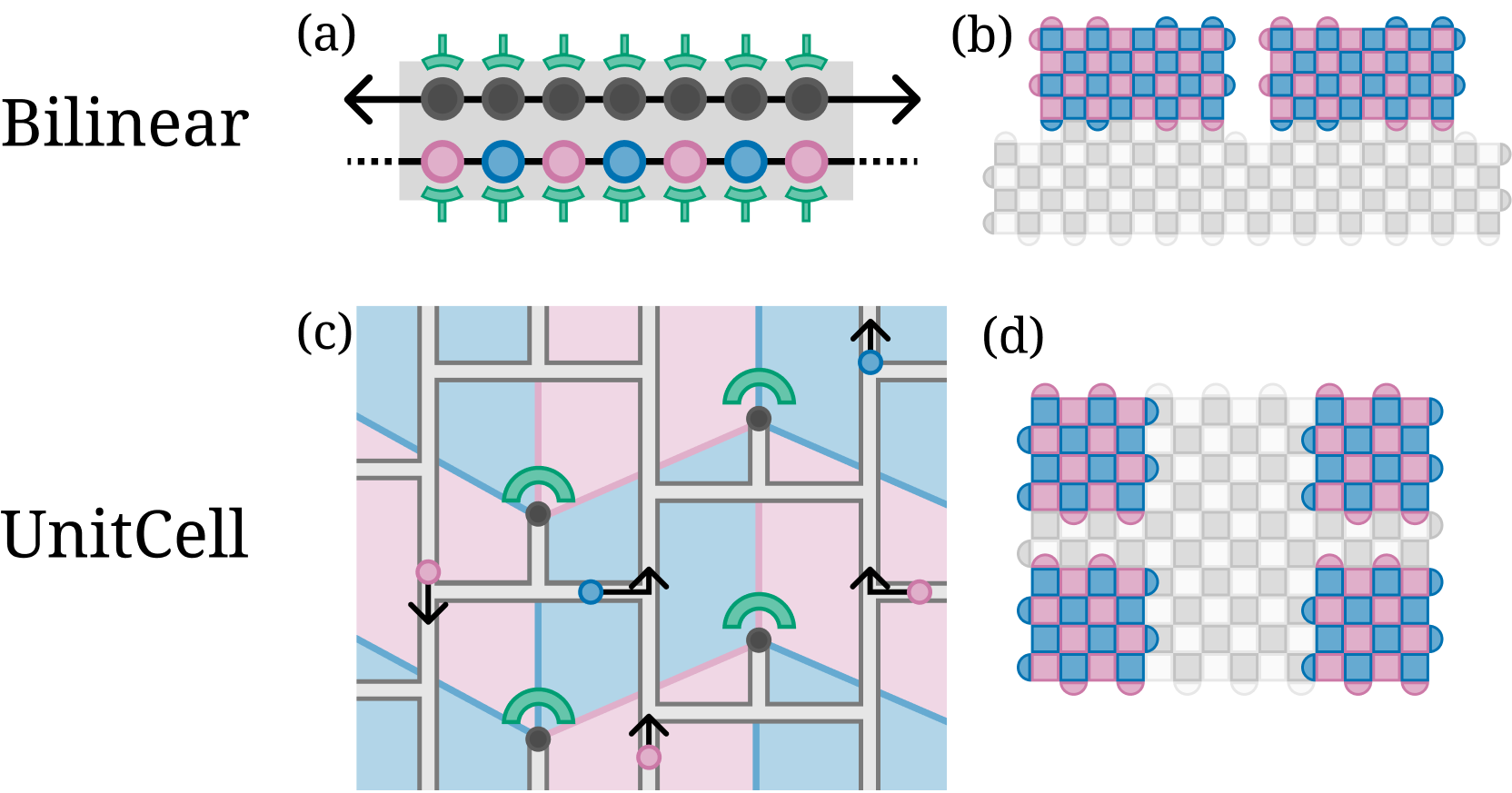}
    \caption{Two prior spin qubit architecture proposals. (a) The \twobyN{} architecture, based on Refs. \cite{siegel_early_2024, siegel_snakes_2025}, consists of two rails of qubits, one of which can shuttle back and forth. Data and ancilla qubits of a surface code are placed column-by-column into the linear array. (b) The \twobyN{} architecture supports a linear logical-level arrangement, where each (wide) surface code patch is connected to others by a shared lattice surgery routing bus (gray). (c) The \spinbus{} architecture, similar to Refs. \cite{boter_spiderweb_2022, kunne_spinbus_2024, otxoa_spinhex_2025}, consists of large unit cells, each of which contains a readout port. A data qubit can be placed in each unit cell and the ancilla qubits must shuttle long distances between unit cells. (d) The two-dimensional physical qubit array of the \spinbus{} architecture naturally produces a two-dimensional logical array, with lattice surgery routing space between surface code patches.}
    \label{fig:baselines}
\end{figure}

We compare SNAQ to two baseline spin qubit architectures based on prior proposals, which we call \spinbus{} and \twobyN{}, \added{depicted} in Figure~\ref{fig:baselines}. These architectures both support a 1:1 ratio of readout ports to qubits, but \added{achieve this by} different means. The \twobyN{} architecture restricts the physical qubits to a 2$\times$N physical layout, as proposed in Ref. \cite{siegel_early_2024}. This design makes fabrication relatively straightforward and cost-effective, but performance depends more heavily on shuttling (as not all data-ancilla pairs can be placed near each other) and yields a logical layout with only one shared lattice surgery routing bus, limiting achievable logical parallelism.

\begin{table}
    \centering
    \begin{tabular}{|l|l|l|l|}
    \hline
         & \spinbus{} & \twobyN{} & \textbf{\snaq{}} (this work) \\
        \hline
        Near-term & \color{purple} $\times$ & \color{ForestGreen} \checkmark & \color{ForestGreen} \checkmark \\
        \hline
        Low-$d$ perf. & \color{purple} $\times$ & \color{ForestGreen} \checkmark & \color{ForestGreen} \checkmark\checkmark \\
        \hline
        Ultrahigh-$d$ perf. & \color{ForestGreen} \checkmark & \color{purple} $\times$ & \color{purple} $\times$ \\
        \hline
        Logical operations & LS & LS & \color{ForestGreen} LS + tCNOT \\
        \hline
        \makecell[l]{Logical clock \\ cycle time ($d=11$)} & 28.6~\si{\micro\second} & 20.8~\si{\micro\second} & \makecell[l]{{\color{ForestGreen} 2.5~\si{\micro\second}} (local) \\ - {\color{purple} 55.6~\si{\micro\second}} (global)} \\
        \hline
        Logical parallelism & \color{ForestGreen} 2D & \color{purple} 1D LS & \color{ForestGreen} 1D transversal \\
        \hline
    \end{tabular}
    \vspace{0.5ex}
    \caption{Comparison to prior silicon architecture proposals}
    \label{tab:baselines}
\end{table}

\removed{On the other hand, t}\added{T}he approach of the \spinbus{} architecture is to create large unit cells that each contain a readout component and have enough space for all readout and qubit wiring, as proposed in Refs. \cite{boter_spiderweb_2022, kunne_spinbus_2024, otxoa_spinhex_2025}. Each unit cell is surrounded by shuttling channels, allowing qubits to be moved between nearby unit cells. We assume a relatively optimistic unit cell size of $5 \times 5$~\si{\micro\meter}, corresponding to 50 dot-to-dot shuttling distances per side.  The surface code can be implemented by assigning each data qubit to its own unit cell and shuttling ancilla qubits between data cells. The benefit of this approach is that the amount of required shuttling does not change with the code distance, so this architecture \removed{is the only one of the three to exhibit a true threshold and to enable}\added{can in principle support} arbitrarily high code distance. The downside is a very large constant amount of shuttling, which can be thought of as effectively adding to the baseline physical error rate, thereby shifting the code threshold. This\removed{will} reduce\added{s} \spinbus{}'s relative performance for smaller code distances. The resulting logical layout is\removed{also} two-dimensional, as shown in Figure~\ref{fig:baselines}d\added{, which allows for improved logical parallelism compared to \twobyN{}}. 

\snaq{}'s layout lies somewhat in between these two baselines with its fixed-width design, though fundamentally differs from both in purposefully deviating from the 1:1 readout-to-qubit ratio of the baselines. This allows \snaq{} to have a much higher qubit density than either baseline, which is critical to enable its transversal logic capabilities.

We provide a high-level, qualitative comparison of these architectural tradeoffs in Table~\ref{tab:baselines}. This summary \removed{contrasts the \spinbus{} and \twobyN{} baselines with \snaq{}, highlighting}\added{highlights} our \removed{proposal's }focus on near-term manufacturability, strong low-distance performance, and efficient logical operations. Unlike the baselines, which are limited to lattice surgery (LS), \snaq{}'s dense design supports fast transversal CNOTs (tCNOTs), enabling a more flexible, high-parallelism logical \added{computation}\removed{layout}.\removed{ The ``1D++'' logical connectivity of \snaq{} refers to the potential parallelism of transversal operations compared to lattice surgery.}

\added{We note that the \twobyN{} architecture proposal has been extended to a two-dimensional architecture supporting transversal operations \cite{siegel_snakes_2025}, but this comes at the cost of increased shuttling, significant added scheduling complexity, and extra qubit costs required to avoid potentially defective shuttling paths. The notion of ``separation distance'' between two logical qubits, which we use to evaluate \snaq{}'s tCNOT, is less well-defined for this architecture, as the logical qubits do not live in a sea of qubits but rather continually move on sparse shuttling tracks. We therefore cannot perform direct comparisons to \snaq{} like we did for \twobyN{} and \spinbus{}. The logical memory performance will be similar to \twobyN{}, and we expect the tCNOT performance to vary significantly with specific hardware parameters like the sparsity of the layout and the overlap between adjacent loops}.

\added{Several proposals have been made for spin qubit architectures based on ``crossbar'' wiring layouts, which offer compelling reductions in wiring complexity \cite{veldhorst_silicon_2017, li_crossbar_2018}. However, in the absence of near-perfect fabrication uniformity across the chip, parallel operation of the physical qubits becomes impossible, leading to severe serialization overhead \cite{li_crossbar_2018, helsen_quantum_2018} that Ref. \cite{pataki_compiling_2025} estimated would limit the best achievable logical error rates to $2 \times 10^{-5}$ (at code distance $d=63$) even with a relatively high coherence time of $10$~\si{\micro\second}. Integrating readout components also remains challenging, with current physical crossbar implementations relegating readout to the edges of the array similar to the \snaq{} arrays considered here \cite{borsoi_shared_2024}.}

\section{Performance evaluations}\label{sec:simulations}

We develop a custom simulation framework that can accurately model the pipelined schedule of shuttled waves of ancilla qubits, allowing us to precisely quantify both the clock speed of the code and the logical performance. We simulate the performance of our proposed architecture using Stim \cite{gidney_stim_2021} and decode with PyMatching \cite{higgott_pymatching_2021}. For a given code distance $d$, we assume a fixed array width of $d+2$. For all experiments, we perform both X and Z-basis experiments and combine the two error rates to obtain an overall error rate.

\subsection{Noise model}

\begin{figure}
    \centering
    \includegraphics[width=\linewidth]{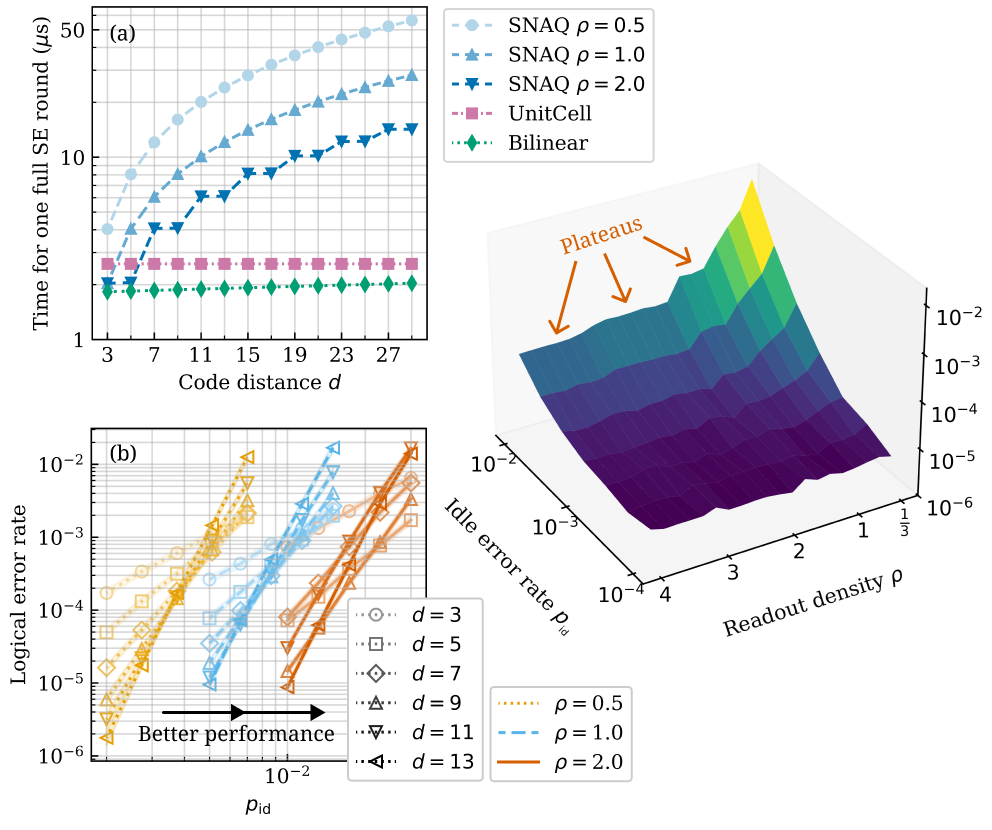}
    \caption{Exploring the effect of readout serialization in \snaq{}. (a) Time to complete one (non-pipelined) SE round in \snaq{} compared to the baselines, for various values of $\rho$. (b) Logical error rate as a function of $p_\text{id}$ for various code distances and densities. Doubling $\rho$ can give an order of magnitude improvement in logical error rate. (c) Logical error rate of a $d=7$ surface code on \snaq{}, showing ``plateaus'' where ranges of density settings lead to the same number of ancilla waves, yielding the same performance.}
    \label{fig:serialization}
\end{figure}

Our noise model is inspired by exchange-only spin qubits \cite{weinstein_universal_2023}; while this limits the direct application of our results for other spin encodings, our general takeaways should apply for a wide range of possible implementations. We use a noise model with three error parameters $p_\text{g}$, $p_{\text{sh}}$, and $p_{\text{idle}}$. The parameter $p_\text{g}$ sets the strength of gate and readout errors. Each CNOT gate causes a two-qubit depolarizing error with probability $p_\text{g}$, each readout encounters a bit flip with probability $p_\text{g}$, and each Hadamard gate causes a one-qubit depolarizing error with probability $p_\text{g}/10$. The parameter $p_\text{sh}$ is the error-per-shuttle such that shuttling a qubit over a distance of $m$ dots incurs a depolarizing error with probability $m \cdot p_\text{sh}$. Finally, the idle error $p_\text{idle}$ is defined as the chance that a qubit experiences a depolarizing idle error in a 1~\si{\micro\second} idle interval (in our noise model, this is equivalent to the time for the five layers of CNOTs in one SE round), so a coherence timescale of $T$ would correspond to $p_\text{id} = e^{-(1~\text{\si{\micro\second}})/T}$.

We set the durations of physical operations assuming an exchange pulse duration of 10~\si{\nano\second}, which means that a CNOT takes approximately 200~\si{\nano\second} (the exact CNOT latency depends on dot-level connectivity \cite{setiawan_robust_2014, chadwick_short_2025}, but here we leave it fixed), a single-qubit Hadamard gate takes 30~\si{\nano\second}, and the shuttling latency is 2~\si{\nano\second} per dot. We assume that initialization and readout can each be done in 500~\si{\nano\second}. Although our numerical simulation results depend on these specific choices, the important underlying ratio is the idle coherence timescale relative to the rate of syndrome extraction rounds. Our results can therefore be interpreted for different hardware latency assumptions by rescaling $p_\text{id}$ accordingly.

In some qubit encodings, the shuttling and idling errors can be strongly biased towards dephasing noise \cite{smet_highfidelity_2024}, which would incentivize asymmetric surface codes with $d_z \neq d_x$. A benefit of \snaq{} is that the readout serialization would be determined by the smaller of $d_x$ or $d_z$. Other options to consider include bias-tailored variants of the surface code \cite{wen_quantum_2003, kovalev_design_2011, terhal_majorana_2012, bonillaataides_xzzx_2021}.

\subsection{Impact of readout serialization}

The readout density $\rho$ is a critical architectural parameter that directly affects logical performance by changing the duration of each syndrome extraction (SE) round, as shown in Figure~\ref{fig:serialization}(a). For $\rho=2$, we see that the SE duration remains below 10~\si{\micro\second} up to $d=21$. Compared to the two baseline architectures, \snaq{}'s SE duration at the same code distance is still significantly longer even with a relatively high $\rho$; however, we will show later that \snaq{} can still achieve a faster logical clock cycle due to its ability to execute transversal gates. \added{Note that pipelining (Figure~\ref{fig:pipelining}) will reduce \snaq{}'s effective SE latency by half.}

To quantify the impact of readout serialization on \snaq{}'s logical performance, we simulate logical \added{memory} performance at different readout densities $\rho$ and different idle error rates $p_\text{id}$, with $p_\text{g} = p_\text{sh} = 0$. Due to the increase in serialization for larger distances, we do not expect to observe a clear threshold (a point where the lines of the same group would all cross). \added{The results are} shown in Figure~\ref{fig:serialization}(b), \added{where we see that} doubling the density can improve the logical performance at the same distance by more than an order of magnitude, underscoring the importance of $\rho$ for the \snaq{} architecture. Figure~\ref{fig:serialization}(c) visualizes this relationship for fixed $d=7$, showing performance ``plateaus'' where ranges of $\rho$ yield the same number of ancilla waves.

\subsection{Comparison to other architectures}\label{sec:memory-comparison}



Although we cannot directly simulate code performance at larger distances with current Monte Carlo methods, as the number of required shots becomes impractical, we can instead fit \added{simulated} lower-distance logical error rates to the expected scaling behavior of \snaq{}. With $p_\text{g}$, $p_\text{sh}$, and $p_\text{id}$ fixed, we can model the \added{idling} logical error rate of \snaq{} as a modification of Eq.~\ref{eq:pL}:
\begin{equation}
    p_L(d) = A\bigg(\frac{p_\text{g}}{p_\text{g}^*} + \frac{p_\text{sh}^\text{eff}}{p_\text{sh}^*} + \frac{p_\text{id}^\text{eff}}{p_\text{id}^*}\bigg)^{(d+1)/2}
\end{equation}
where $p_\text{sh}^\text{eff}$ is the cumulative shuttling error experienced by the physical qubits \added{and} $p_\text{id}^{\text{eff}}$ is the cumulative idling error experienced by the qubits. Here we have made the first-order assumption of a linear threshold surface \cite{hetenyi_tailoring_2024} characterized by three thresholds $p_\text{g}^*$, $p_\text{sh}^*$, and $p_\text{id}^*$. To fit circuit-level simulation data to this model, we use the form
\begin{equation}
    p_L(d) = A\big(\alpha +\beta d + \gamma n_w(\rho, d)\big)^{(d+1)/2},
    \label{eq:fit}
\end{equation}
where $A$, $\alpha$, $\beta$, and $\gamma$ are fitting parameters, and $n_w(\rho, d)$ is given by Eq.~\ref{eq:nw}. This model accounts for the three ways in which physical errors in \snaq{} scale with increasing code distance: gate errors remain constant, shuttling errors increase linearly in $d$, and idle errors increase with the amount of ancilla serialization $n_w$. We have derived this model from the structure of the SE circuit, and we find that it fits well to the available simulation data, but we caution that it is an approximation and circuit-level simulation still remains the most reliable performance predictor. To fit the \twobyN{} baseline, we fix $\gamma=0$, and for the \spinbus{} baseline, we fix $\beta=\gamma=0$.

\begin{figure}
    \centering
    \includegraphics[width=\linewidth]{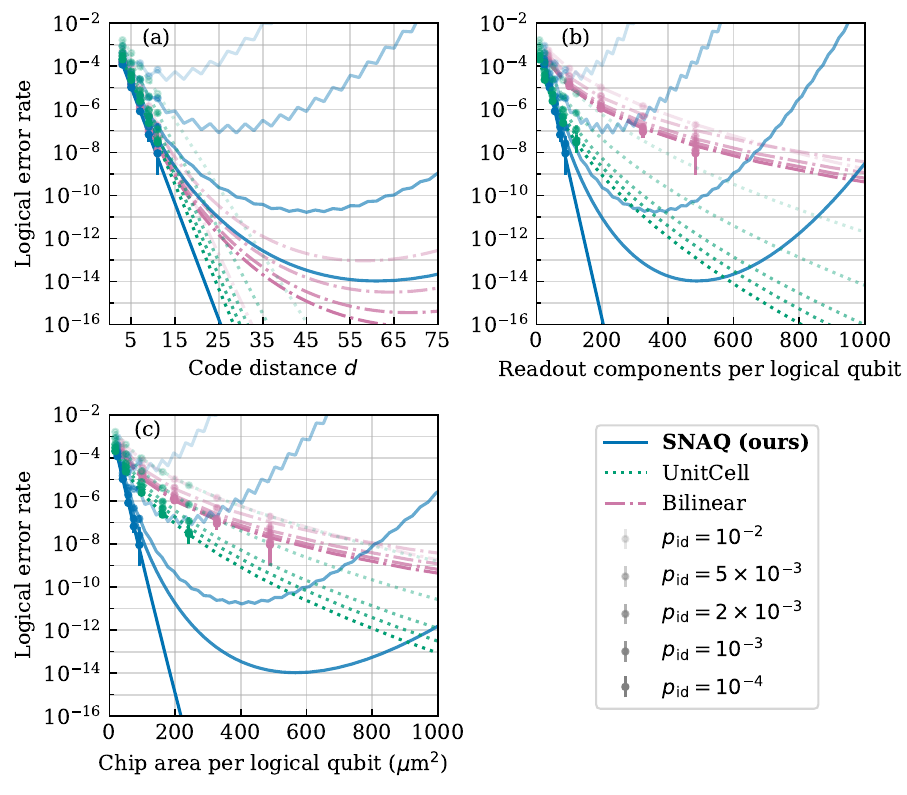}
    \caption{Projecting logical performance to higher code distances under fixed $(p_\text{g}, p_\text{sh}) = (10^{-3}, 10^{-5})$ for various $p_\text{id}$ with \snaq{} $\rho=2$. (a) \snaq{} and \twobyN{} give diminishing returns as code distance increases due to an increase in shuttling and idling errors with larger code block size, while \spinbus{} exhibits consistent error suppression as distance is increased. (b) Logical error rate as a function of readout count, which may be an important driver of packaging cost, showing that \snaq{} outperforms \twobyN{} and is competitive with \spinbus{}. (c) Logical error rate as a function of total component area per logical qubit. For sufficiently low $p_\text{id}$, \snaq{} achieves better logical performance at significantly reduced total chip area compared to both baselines.}
    \label{fig:extrapolating_id}
\end{figure}

In Figure~\ref{fig:extrapolating_id}(a), we simulate the performance of \snaq{} and baselines for distances up to $d=11$ under $(\rho, p_\text{g}, p_\text{sh}) = (2, 10^{-3}, 10^{-5})$ and various values of \added{the critical parameter} $p_\text{id}$, fitting each set of results to Eq.~\ref{eq:fit} and extrapolating to large distances. \removed{As expected, \snaq{} is highly sensitive to $p_\text{id}$, but we see that it can still reach utility-scale error rates below $10^{-6}$ with $p_\text{id}=5 \times 10^{-3}$, roughly corresponding to a coherence time of 200~\si{\micro\second}, which is well within achievable ranges for spin qubits. This figure shows that a}\added{A}t a fixed code distance, \snaq{}'s logical error rate is far more sensitive to $p_\text{id}$ than the baselines. \added{As expected,} we\removed{also} can clearly see that $p_\text{id}$ defines a\added{n}\removed{clear} error floor for \snaq{}, limiting the achievable logical error rate. \added{However, we see that it can still reach utility-scale error rates below $10^{-6}$ with $p_\text{id}=5 \times 10^{-3}$, roughly corresponding to a coherence time of 200~\si{\micro\second}, which is well within achievable ranges for spin qubits.} To reach extremely low error rates below $10^{-12}$ for long-term resource-intensive applications, $p_\text{id}=10^{-3}$ or $10^{-4}$ will be required. This translates to a spin qubit coherence time near the single-\si{\milli\second} range. Although this has already been demonstrated in small devices \cite{muhonen_storing_2014, veldhorst_addressable_2014, malinowski_notch_2017, laucht_dressed_2017}, it remains to be seen whether the same performance can be achieved at scale. \added{Reductions in gate and measurement duration compared to the assumptions in this work would relax these coherence time targets.}

However, comparing these architectures at a fixed code distance is misleading because it obscures the starkly different physical resource costs of the architectures. In \snaq{}, where I/O costs and qubit count are decoupled, the number of qubits may no longer be the primary cost driver. We therefore turn to more relevant architectural metrics by studying two fabrication-limited resources: the total number of readout components\added{, which is a direct driver of I/O complexity,} and the\removed{resulting} chip area \added{per logical qubit}. Figure~\ref{fig:extrapolating_id}(b) shows the same \added{simulation} data but with code distance converted to the number of readout components required per logical qubit, revealing \snaq{}'s much more efficient use of readout components. Finally, in Figure~\ref{fig:extrapolating_id}(c), we convert code distance to chip area by assuming that each \added{physical} qubit (plunger and adjacent barrier gates) takes up a $100\times 100$~\si{\nano\meter} $ = 0.01$~\si{\micro\meter}$^2$ space and\removed{we use} \added{using} an order-of-magnitude estimate of the readout footprint as $1$~\si{\micro\meter}$^2$ based on Ref. \cite{schmidt_compact_2024} and chip images from e.g. Refs. \cite{morello_singleshot_2010, connors_rapid_2020, hendrickx_singlehole_2020, curry_singleshot_2019, oakes_fast_2023}. These estimates account for the footprint of the components themselves, but do not include interconnect routing, which will be implementation-specific. For \spinbus{}, we therefore only consider the area taken up by the readout component and shuttling channels (not the large interior spaces reserved for wiring). This analysis reveals that, for achievable $p_\text{id}$, \snaq{} produces logical qubits that are orders of magnitude more area-efficient than the baselines, already providing a clear advantage at $p_\text{id}=5\times 10^{-3}$ that becomes much stronger with lower $p_\text{id}$.

\subsection{Logical operations in \snaq{}}\label{sec:tcnot-fid}

\added{An important difference between lattice surgery and transversal CNOTs in \snaq{} is the scaling of their error rates as the separation between the two surface code patches increases:}\removed{However, the separation distances at which a standard tCNOT will outperform lattice surgery are limited:} in \snaq{}, the \textit{physical} shuttling and idling errors that occur during a transversal CNOT will increase linearly with the distance between the surface code patches, meaning that the \textit{logical} error rate will increase\removed{exponentially} \added{polynomially with degree $(d+1)/2$ once the shuttling errors become dominant, \added{which implies} that code distance cannot be increased indefinitely to extend the tCNOT's range. In contrast, lattice surgery errors increase linearly with the total merge volume due to a linearly increasing number of possible error chains, so it is always possible to increase the code distance to enable further-distance communication}. We thus expect transversal operations to perform better for sufficiently close patches, while lattice surgery will be preferred for sufficiently long-distance communication.

\begin{figure}
    \centering
    \includegraphics[width=0.9\linewidth]{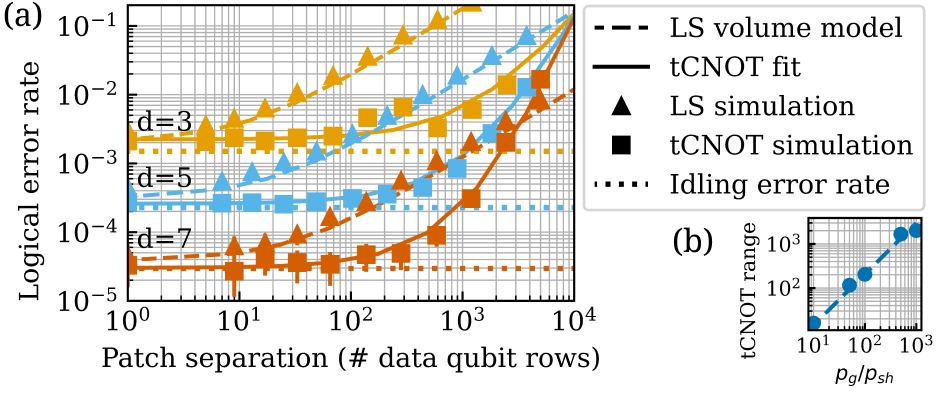}
    \caption{\added{(a)} Error rates of transversal CNOT (tCNOT) and lattice surgery merge-split (LS) as a function of patch separation, which is defined as the number of data qubit rows between the two patches. \added{(b) The range in which tCNOT incurs comparable error to idling is limited by the accumulation of shuttling errors, and therefore scales linearly with the relative strength of $p_\text{sh}$.}}
    \label{fig:tcnot_vs_ls}
\end{figure}

To investigate this tradeoff, we perform circuit-level simulations of tCNOT and LS operations in the \snaq{} architecture using parameters $(p_\text{g}, p_\text{id}, p_\text{sh}, \rho) = (10^{-3}, 10^{-4}, 10^{-5}, 1.0)$. We prepare two surface code patches in the logical $\ket{0}$ state and either perform a tCNOT or an LS merge-split \added{XX measurement} between them, calculating the resulting error rate\added{, which is} \removed{(}the chance that either logical qubit experiences a bit flip\removed{ for the tCNOT, and the chance that we correctly measure the ZZ observable for the LS operation)}. \added{We include $d$ rounds of idling on each patch before and after the operation. We compare the error rate of this experiment to that of two patches simply idling for $2d$ rounds.} We decode the lattice surgery experiment with PyMatching and the transversal CNOT with BP+OSD \cite{roffe_decoding_2020, roffe_ldpc_2022}. The results are shown in Figure~\ref{fig:tcnot_vs_ls}\added{(a): the error rate of distance-$d$ lattice surgery scales linearly with increasing separation distance, while the tCNOT's error rate remains low (comparable to the idling surface code) until accumulated shuttling errors become stronger than the gate error $p_\text{g}$, at which point the error rate increases polynomially. The tCNOT's limit is specified by the relative strength of $p_\text{sh}$ to other error mechanisms, which we demonstrate in Figure~\ref{fig:tcnot_vs_ls}(b) by calculating the $d=5$ tCNOT's effective range for different ratios of $p_\text{g}/p_\text{sh}$. We define the range as the largest separation distance such that the tCNOT adds a negligible (within sampling uncertainty) additional error compared to the idling experiment.}\removed{, where we see that tCNOT and LS both exhibit performance near memory idle error rates for short separation distances, but for sufficiently large distances the performance of tCNOT begins to degrade. The distance at which this occurs depends on the relative strength of the per-dot shuttling error compared to other error sources: performance suffers once the separation distance $s$ is large enough that $s\cdot p_\text{sh}$ is the dominant source of error. In the case of our simulations, $p_\text{g} = 10^{-3}$ is the dominant error source for an idling surface code, so with $p_\text{sh}=10^{-5}$, the limit will be around $s\approx100$, which matches what we observe in the simulation results.}

\added{We find that the LS error rate in Figure~\ref{fig:tcnot_vs_ls}(a) is well-modeled by a linearly scaling increase in logical error rate $p_L(s) = \frac{(s+d)d^2}{4d^3} \cdot p_L(1) + p_L^{\text{idle}}$ for a separation distance of $s$ data qubit rows, where the prefactor gives the relative spacetime volume of the merge operation, $p_L(1)$ is the lattice surgery error rate at minimal separation distance, and $p_L^\text{idle}$ is the idling logical error rate. We find that the tCNOT's error rate is well-modeled by $p_L = A (B(s+d) + C)^{(d+1)/2}$ after fitting $A$, $B$, and $C$, revealing the expected polynomial scaling behavior once $s$ is sufficiently large.}

\begin{figure}
    \centering
    \includegraphics[width=0.9\linewidth]{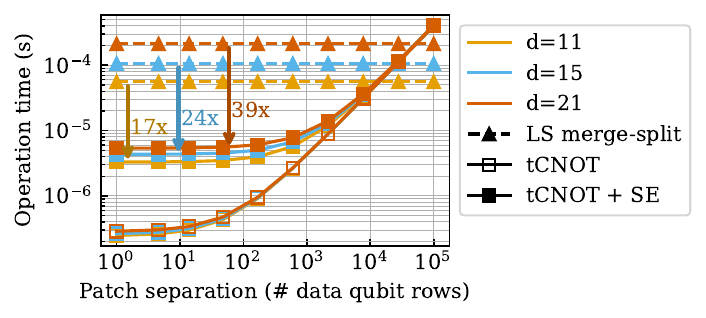}
    \caption{Duration of lattice surgery merge-split and tCNOT operations in \snaq{} as a function of patch separation, assuming a per-shuttle latency of 2~\si{\nano\second} and a density of $\rho=1$. \added{tCNOT+SE is the latency for one tCNOT and half an SE round, which is the effective operation latency in a compiled program.} For separation distances under 200 dots, tCNOT takes less than a microsecond (without the SE round), and tCNOT+SE is competitive with lattice surgery out to separation distances over 10,000.}
    \label{fig:tcnot_vs_ls_time}
\end{figure}

We envision several potential ways to extend the tCNOT to longer \added{effective} ranges, all of which involve a space tradeoff in the \snaq{} array. First, the shuttled surface code could stop along the way to perform multiple SE rounds during a long-distance tCNOT, which would avoid accumulating shuttle errors \cite{bravyi_highthreshold_2024}. Second, extra spaces could be used to prepare Bell pairs to use for gate teleportation or entanglement swapping \cite{fowler_surface_2010}. Finally, the most aggressive method is to designate every other surface code as a dedicated logical ancilla for GHZ state preparation, allowing for constant-depth CNOTs \cite{yang_harnessing_2024, baumer_measurementbased_2025, rines_demonstration_2025}. \removed{Although these methods are compelling, an }\added{These methods may enable long-distance communication at significantly reduced cost compared to lattice surgery; however, a thorough} evaluation of the tradeoffs involved is beyond the scope of this work.

\subsection{Logical clock speed}

\added{To further compare the two modes of logical operation in \snaq{}, we calculate the latencies of the operations, which directly determine the logical clock speed of the fault-tolerant processor.} Figure~\ref{fig:tcnot_vs_ls_time} shows the durations of the tCNOT and LS merge-split operations \added{in \snaq{}} over varying separation \added{distance} for three surface code distances $d \in (11, 15, 21)$. The LS latency is the time to complete $d$ pipelined SE rounds. For the tCNOT, we show both the time to complete the tCNOT itself \removed{(consisting of shuttling and a physical CNOT layer) and the}\added{as well as the overall average operation latency}\removed{time to complete a tCNOT and half a non-pipelined SE round,} assuming that two tCNOTs are performed for every SE round \cite{cain_correlated_2024, zhou_resource_2025}. \added{We can speed up these individual SE rounds by using both edges of the array for initialization and measurement of the ancilla qubits, roughly halving the number of serialization waves. For separation distances within the tCNOT's fidelity-limited range of 100, the tCNOT+SE latency is effectively constant and is consistently over 10$\times$ faster than LS}\removed{ For separation distances up to $10^4$, we observe that the tCNOT+SE latency is significantly lower than the LS operation by up to 10$\times$ depending on the code distance. The LS operation slightly increases in duration for larger separation distances due to the need to preserve performance (by adding SE rounds) against an increasing number of possible temporal error chains \protect\cite{domokos_characterization_2024}}.

\begin{figure}
    \centering
    \includegraphics[width=\linewidth]{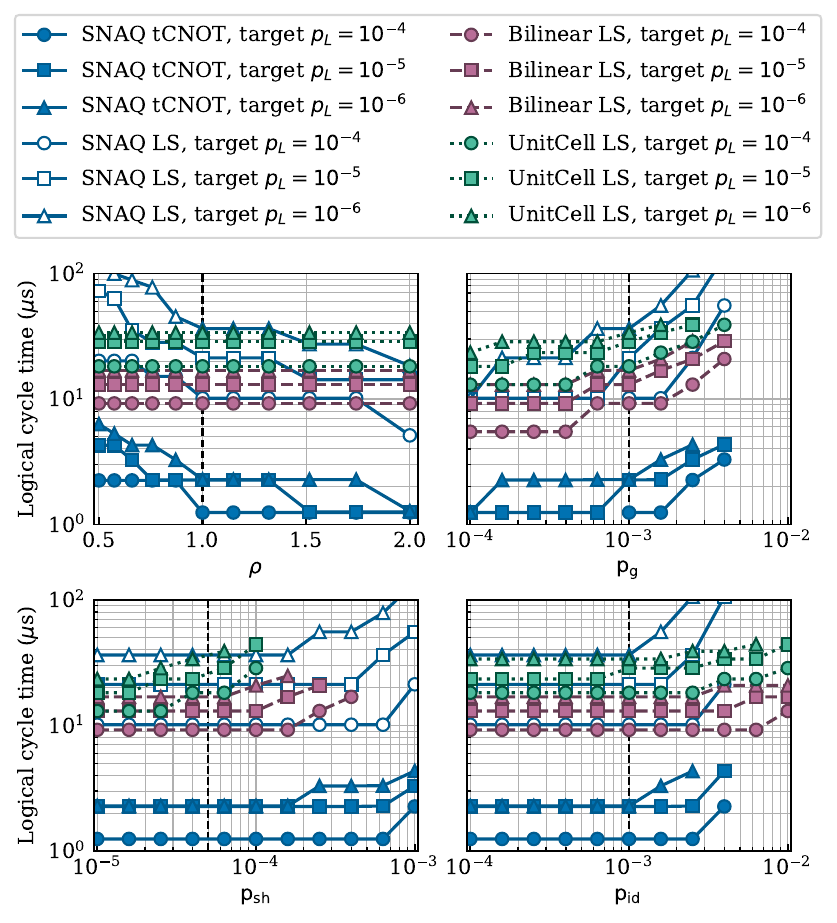}
    \caption{\added{Logical cycle times of proposed spin qubit architectures under varying noise parameters at fixed target logical error rates. Each plot shows sensitivity to one of the four hardware parameters $(\rho, \text{p}_\text{g}, \text{p}_\text{sh}, \text{p}_\text{id})$. Vertical dashed lines indicates default parameter values when held constant in other plots. Points are omitted for parameter settings where no simulated code distance could reach the target logical error rate. \snaq{}'s tCNOT maintains a consistent latency advantage across a wide range of hardware parameters, while \snaq{}'s LS operation is generally slower than that of the baselines. \snaq{} is more resilient to high $\text{p}_\text{sh}$ than the baselines (lower left), but fails for sufficiently-high $\text{p}_\text{id}$ (lower right).}}
    \label{fig:improvements_grid}
\end{figure}

\added{In Figure~\ref{fig:improvements_grid}, we compare logical clock speed to the baseline architectures, varying each hardware parameter across a wide range, for three target logical error rates $p_L \in (10^{-4}, 10^{-5}, 10^{-6})$. For each instance, we simulate code distances up to 15, find the smallest distance that reaches the target $p_L$, and calculate the logical cycle time of the code. \snaq{} tCNOT consistently maintains an advantage across the parameter space, with the exception of high $p_\text{id}$ (lower right), where no instances of \snaq{} could reach the target logical error rate. \snaq{} can operate even with very high shuttling error, where the baselines fail (lower left). This demonstrates the key tradeoff of the \snaq{} architecture compared to the baselines: physical qubits are packed much more closely, reducing required shuttling costs but increasing idling time due to readout serialization. \snaq{}'s improvement factor is most sensitive to the readout density $\rho$ (upper left).}

\begin{figure}
    \centering
    \includegraphics[width=\linewidth]{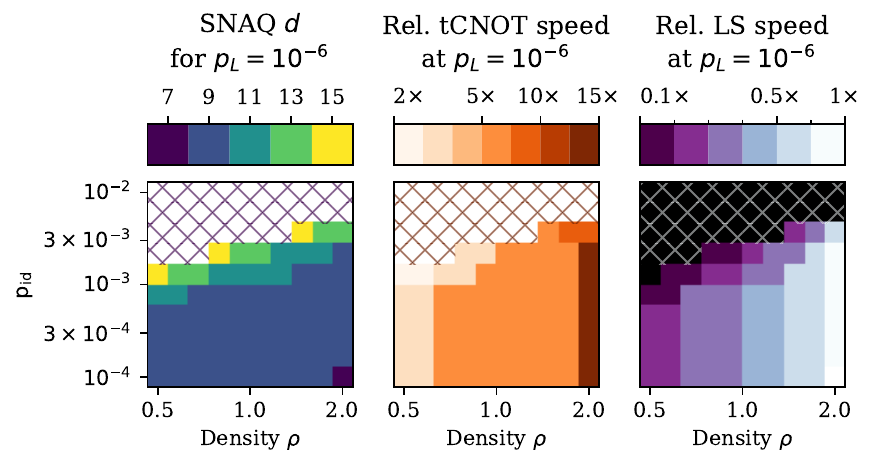}
    \caption{\added{Quantifying \snaq{}'s logical performance as $\rho$ and $\text{p}_\text{id}$ vary. Parameters $\text{p}_\text{sh}$ and $\text{p}_\text{g}$ are held constant at $5 \times 10^{-5}$ and $10^{-3}$, respectively. Hatched area shows parameter range where no \snaq{} instance reached the target logical error rate of $p_L \leq 10^{-6}$. Center plot shows relative speedup of \snaq{}'s tCNOT compared to a lattice surgery operation on the baselines (choosing the faster baseline at each point), demonstrating logical clock speedups between $2 \times$ and $15 \times$. Rightmost plot shows relative speed of \snaq{}'s LS operation compared to the best baseline, showing a significant slowdown up to $10\times$ for low $\rho$ and high $\text{p}_\text{id}$.}}
    \label{fig:improvements_heatmaps}
\end{figure}

\added{To further investigate \snaq{}'s dependence on $\rho$, we study this tradeoff in more depth in Figure~\ref{fig:improvements_heatmaps}. For fixed target $p_L=10^{-6}$ and varying $p_\text{id}$ and $\rho$, we again calculate the required code distance and then convert to logical clock speeds. Across the studied range of $\rho$, we see that \snaq{}'s tCNOT speedup varies from 2$\times$ to $15\times$, and its LS slowdown varies from 10$\times$ to negligible. We also see that \snaq{} fails for $p_\text{id}$ significantly above $10^{-3}$, with the exact failure point depending on $\rho$.}

\begin{figure}
    \centering
    \includegraphics[width=\linewidth]{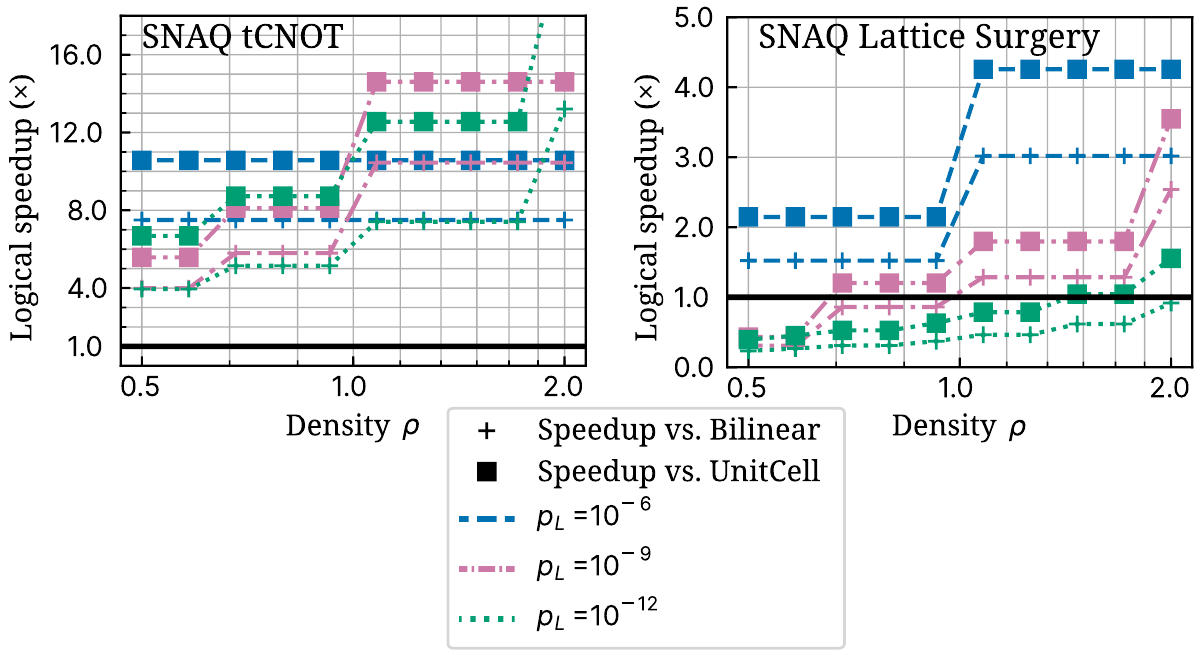}
    \caption{\added{Comparing projected performance at lower logical error rates}\removed{Comparing \snaq{} logical operations to baselines} under $(p_\text{g}, p_\text{sh}, p_\text{id}) = (10^{-3}, 10^{-5}, 10^{-4})$ for various $\rho$\added{, assuming patches are adjacent}. \snaq{}'s tCNOT outperforms baseline lattice surgery methods by over $4\times$ even for low $\rho$ and can provide up to \removed{$14.6\times$}\added{$22\times$} improvement in the range studied here. Lattice surgery on \snaq{} is fast for higher $p_L$ and $\rho$, but becomes less efficient for lower $p_L$ due to the significant increase in required code distance (making each SE round take much longer).}
    \label{fig:speedup_vs_density}
\end{figure}

\removed{To more \removed{thoroughly}\added{precisely} quantify \snaq{}'s potential speedup and \added{to study} its sensitivity to the key hardware parameter $\rho$, }\added{Finally, we consider performance at lower $p_L$ by extrapolating the simulated data using the models discussed in Section~\ref{sec:memory-comparison}.}\removed{We simulate logical performance for various values of $\rho$.} For this analysis, we use fixed physical error rates of $(p_\text{g}, p_\text{sh}, p_\text{id}) = (10^{-3}, 10^{-5}, 10^{-4})$, simulate distances up to 11, and fit the model in Eq.~\ref{eq:fit} to the data \added{for a range of values of $\rho$}. \removed{We apply the same process to the two baselines to provide a direct comparison. As in Section~\ref{sec:memory-comparison}, t}These fits allow us to determine the code distance \removed{at which}\added{needed for} each architecture \removed{can}\added{to} achieve a \added{certain}\removed{given} logical error rate\removed{. We}\added{, which we} can then convert\removed{these code distances } to logical clock speeds. Figure~\ref{fig:speedup_vs_density} shows the comparison of the logical operation speeds in \snaq{} to those of \spinbus{} and \twobyN{}. We find that \snaq{}'s tCNOT is significantly faster than either of the baselines across the entire studied range. For a near-term target \removed{$p_L$}\added{error rate} of $10^{-6}$, we find consistent 7.5$\times$ and 10.6$\times$ speedup\added{s} across the density range. For lower $p_L$, the improvement is smaller\removed{ for low $\rho$,} but still remains above 4$\times$. Because the interaction distance of the tCNOT is limited\removed{, as we will investigate in the following subsection}, we also compare the lattice surgery time between \snaq{} and the baselines\removed{. Additionally, we find}\added{, finding} that \snaq{}'s lattice surgery still outperforms the baselines for a target $p_L$ of $10^{-6}$ but is less competitive for lower $p_L$ and lower $\rho$.

\subsection{Architectural implications}

The \added{fidelity and latency}\removed{ speed and error rate} studies discussed above suggest that both tCNOTs and lattice surgery are valid options for logical operations on the \snaq{} architecture, with tCNOTs providing a significant speed advantage for sufficiently short separation distance and LS operations enabling longer-distance communication without degrading fidelity\added{, but at the cost of over 10$\times$ slower operation speed.}\removed{. However, the LS operations are approximately 10$\times$ slower that the tCNOT (depending on patch separation and code distance).}

A \snaq{} device of width $w$ can support a surface code distance of up to $d=w-2$ in a logical $1\times N$ configuration if logical operations are restricted to only tCNOTs. This processor would have \added{a}\removed{some} maximum allowed interaction distance within which the tCNOT's performance is acceptable, which \added{would} depend\removed{s} on the relative strengths of shuttling errors to other sources of error. \removed{Compiling programs to such an architecture}\added{This limited connectivity} is reminiscent of\removed{the challenges faced in} nearest-neighbor-connect\added{ed}\removed{ivity} noisy intermediate-scale quantum (NISQ) processors, where long-range interactions require multi-hop, higher-cost operations, except in this case the extra cost is in \removed{time}\added{latency} instead of added error. In these NISQ processors, locality-aware mapping and routing techniques were developed to minimize the amount of long-distance communication needed \cite{li_tackling_2019}; \snaq{} may benefit from the \added{adaptation} of \added{these} techniques to make use of fast tCNOTs whenever possible. \added{Compiling to a transversal-equipped architecture presents several interesting opportunities to further improve processor speed. First, logical shuttles in the same direction can be performed in parallel. Second, only one of the two logical qubits is actively in-use during the shuttling periods of a tCNOT, allowing the other qubit to take part in other operations in the meantime. Third, the end-to-end latency of a longer-distance tCNOT can be nearly halved if the mobile qubit is not shuttled back to its original location, but instead deposited close to the stationary qubit after the operation.}

On the other hand, \removed{a}\added{the same width-$w$} \snaq{} device\removed{of width $w$} could also support a surface code distance of up to $d=\lfloor \frac{w-3}{2} \rfloor$ in a logical $2\times N$ layout, \removed{which could then support}\added{enabling} both tCNOTs and lattice surgery if one logical channel is left free to use as routing space. A processor with two available modes of communication is reminiscent of the latency hierarchies in classical computer architecture, where care must be taken to minimize the use of main memory accesses or network connections. This presents a compelling challenge for future compilation research.

\section{Fault-tolerant resource estimation}

\begin{figure*}
    \centering
    \includegraphics[width=0.8\linewidth]{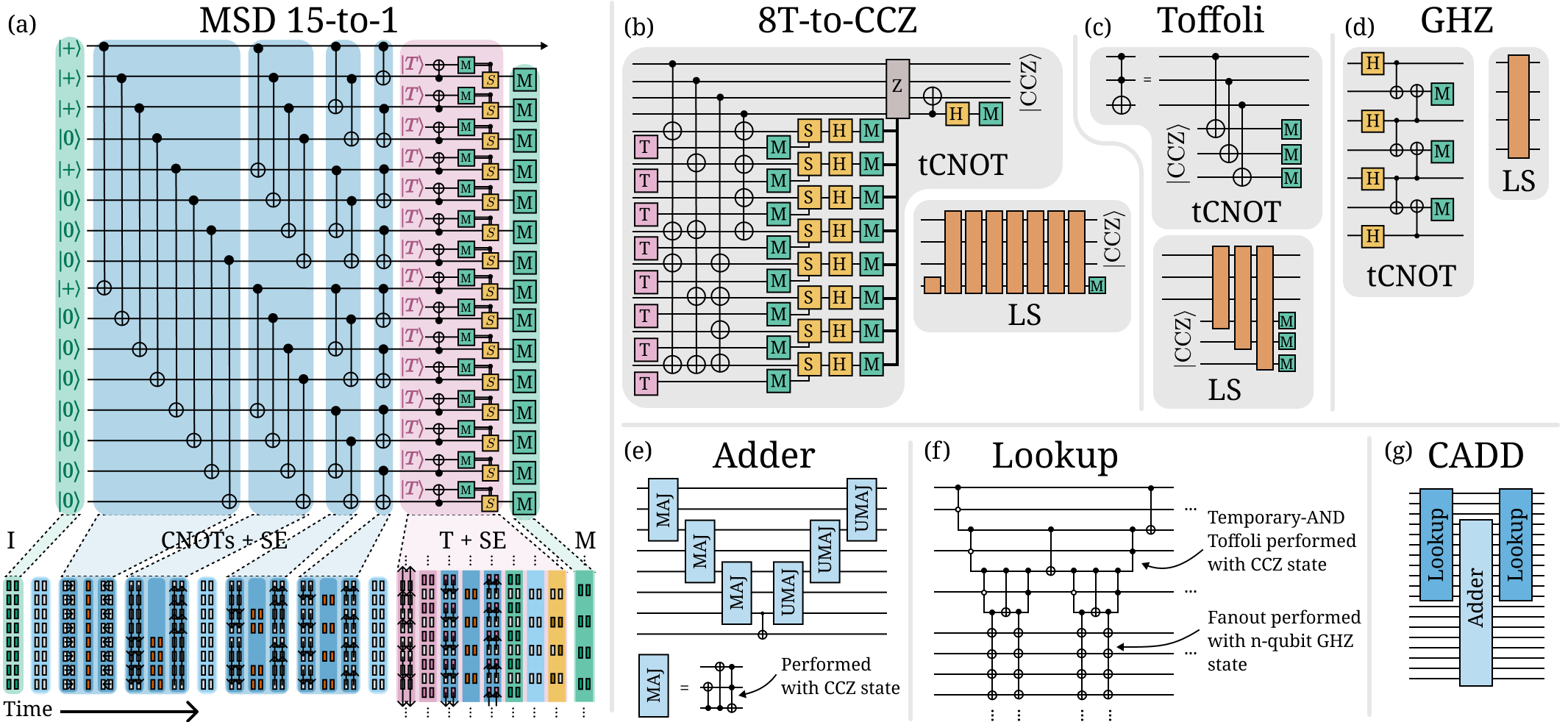}
    \caption{\removed{\textit{Top:}}\added{(a)} Logical circuit implementing 15-to-1 T distillation. CNOTs are shaded in groups to show those that can be performed in parallel using transversal CNOTs. \textit{Bottom:} A $2 \times 8$ \snaq{} layout executing the circuit. The schedule involves parallel tCNOTs and T gate injection, interleaved with four SE rounds. \added{(b-d) Logical primitives described in Section~\ref{sec:primitives}, expressed in CNOT and lattice surgery (LS) circuits: 8T-to-CCZ distillation, a Toffoli gate (note: auto-corrections not shown), and GHZ state preparation. (e-f) Larger logical subroutines described in Section~\ref{sec:subroutines}: Adder \cite{cuccaro_new_2004} and Lookup \cite{zhou_resource_2025}. (g) The lookup-controlled adder (CADD) is the key component of the quantum factoring algorithm.}}
    \label{fig:benchmarks}
\end{figure*}

The numerics in the prior section showed that, using transversal CNOTs, the \snaq{} architecture can perform individual logical operations \added{2-15$\times$}\removed{over twice} as fast as competing architectures for sufficiently close logical qubits. \added{Such close-range tCNOTs support key application subroutines including quantum arithmetics and state preparation, which in turn can implement important quantum applications such as factoring and chemistry simulation\cite{zhou_resource_2025}. In this section, we explicitly compile several key primitives---T and CCZ distillation, the Toffoli gate, GHZ state generation, and long-range transport---and use them to calculate the costs of larger subroutines (Adder and Lookup), with fixed $\rho=1$ and $d=27$. With these subroutines as building blocks, we heuristically estimate the improvements SNAQ can achieve compared against baselines. Although large scale algorithms generally be nonlocal, we find that the qubit transport costs are sufficiently low that the tCNOTs still provide a speed advantage.
In particular, we estimate up to  $2.4\times $ and $2.9\times$ reductions over the Bilinear and UnitCell baselines for factoring, and expect significant reductions for various chemistry simulation algorithms as well.}\removed{As concrete example of compiling logical circuit to the \snaq{} architecture, we consider 15-to-1 T state distillation.}

\added{\subsection{Primitives: state preparation, distillation, and logic}\label{sec:primitives}}

We simulate the runtime of several key fault-tolerant circuit primitives by compiling them to either transversal or lattice surgery instruction sets and calculating the execution time for each architecture. For \snaq{}, we insert an SE round after every two layers of tCNOTs. In the following section, we describe the primitives we consider.


\subsubsection{15-to-1 T state distillation}

The 15-to-1 distillation circuit \removed{is derived from the [[15,1,3]] Hastings-Haah code \cite{bravyi_universal_2005} and} involves preparing 15 noisy T states and \removed{``distilling'' them}\added{performing an encoding circuit} to produce one higher-fidelity T state. 
For \snaq{}, we \removed{choose to use the }\added{use a tCNOT-compatible }circuit shown in Figure~\ref{fig:benchmarks}a\removed{ which uses 16 logical qubits. The benefit of this encoding circuit is that the transversal CNOT implementation is highly parallelizable, requiring only four layers if the shuttles are properly scheduled}. We assume \removed{readout density $\rho=1$, }T state preparation time comparable to surface code initialization time\removed{,} and no decoding delay for conditional S corrections. Our construction assumes that four SE rounds are performed during the distillation, as shown in Figure~\ref{fig:benchmarks}.\removed{Each SE round here takes twice as long due to the two-column logical layout.} We assume the implementation of a fold-transversal S gate \cite{moussa_transversal_2016, chen_transversal_2024}, which we set to have a latency equal to $2d\cdot t_\text{shuttle} + t_\text{CNOT}$. Neither of the two prior spin qubit architecture proposals is compatible with parallelizable transversal CNOTs, so lattice surgery is needed instead. For the \spinbus{} architecture, we can use the \added{implementation} from Figure 11 of Ref. \cite{litinski_magic_2019}, which requires $6d$ SE rounds on a 15-patch two-dimensional layout \added{of surface codes}. For the \twobyN{} architecture, using a virtual 2$\times$N layout of logical surface code qubits as depicted in Figure 3 of Ref. \cite{siegel_early_2024}, we can implement the same circuit\removed{, but must} restrict\added{ed} to a 2$\times$5 layout\added{, which extends the temporal cost to $12d$ SE rounds}.

\begin{table*}
    \centering
    \begin{tabular}{|r|r|r|r|r|r|r|r|l|}
        \hline
         & \multicolumn{2}{c|}{MSD 15-to-1} & \multicolumn{2}{c|}{8T-to-CCZ} & \multicolumn{2}{c|}{GHZ, $n=100$} & \multicolumn{2}{c|}{$k$-patch transport}\\
         \cline{2-9}
         & Time (\si{\micro\second}) & Vol. (qubit$\cdot$s) & Time (\si{\micro\second}) & Vol. (q$\cdot$s) & Time (\si{\micro\second}) & Vol. (q$\cdot$s) & Time (\si{\micro\second}) & Vol. (q$\cdot$s) \\
         \hline
         \twobyN{} & 654.5 & 9.536 & 2182.1 & 31.793 & 54.5 & 15.893 & 54.5\hspace{1.2ex} & 0.0795($k$+2) \\
        \spinbus{} & 421.2 & 9.205 & 2808.5 & 40.920 & 70.2 & 20.456 & 39.0\hspace{1.2ex} & 0.102($k$+2) \\
        \textbf{\snaq{}} & \textbf{302.2} & \textbf{7.045} & \textbf{250.0} & \textbf{7.284} & \textbf{18.7} & \textbf{5.426} & 4.0$k$ & \textbf{0.00586$k$} \\
        \hline
    \end{tabular}
    \vspace{0.5ex}
    \caption{Cost comparison of fault-tolerant primitives at $d=27$.}
    \label{tab:primitives}
\end{table*}

        

\begin{table}[htbp]
    \centering
    
    \begin{tabular}{|r|r|r|r|r|}
        \hline
          & \multicolumn{2}{c|}{Adder, $n=139$} & \multicolumn{2}{c|}{Lookup($7, 139$)}\\
         \cline{2-5}
          & Time (\si{\milli\second}) & Vol. (q$\cdot$s) & Time (\si{\milli\second}) & Vol. (q$\cdot$s)\\
         \hline
         \twobyN{} & \textbf{3.767} & 21350.9 & 3.463 & 1535.0\\
        \spinbus{} & 4.002 & 23055.4 & 4.458 & 2014.7\\
        \textbf{\snaq{}} & 7.584 & \textbf{13446.9} & \textbf{0.711} & \textbf{354.4}\\
        \hline
    \end{tabular}
    
    \vspace{2ex} 
    
    \begin{tabular}{|r|r|r|r|r|}
        \hline
          & \multicolumn{2}{c|}{CADD$(10, 1067)$ Ref. \cite{gidney_how_2025}} & \multicolumn{2}{c|}{CADD$(7,139)$ Ref. \cite{zhou_resource_2025}}\\
         \cline{2-5}
          & Time (\si{\milli\second}) & Vol. (q$\cdot$s) & Time (\si{\milli\second}) & Vol. (q$\cdot$s)\\
          \hline
        \twobyN{} & 79.1 & 1838772 & 10.7 & 60918\\
        \spinbus{} & 95.4 & 2225879 & 13.0 & 74821\\
        \textbf{\snaq{}} & \textbf{76.5} & \textbf{756051} & \textbf{10.0} & \textbf{17718}\\
        \hline
    \end{tabular}
    
    \vspace{0.5ex}
    \caption{Full costs of logical subroutines and controlled adders (CADD) at $d=27$. Includes CCZ factories and qubit transport costs. CADD dominates the cost of quantum factoring.}
    \label{tab:merged-subroutines}
\end{table}

\removed{The results are shown in Table~\ref{tab:primitives}, where we see that \snaq{} achieves a \removed{\removed{60-63}\added{58-60}\% volume reduction compared to the baselines at $d=7$ and }\removed{64}\added{57-58}\% volume reduction at $d=15$. These cost reductions, while significant, are smaller than the improvements in the logical clock speed. We attribute this to two aspects of this particular example: (1) the compiled distillation circuit is very shallow, with only five layers of tCNOTs, so placing an SE round \added{after initialization and} after every two tCNOTs yields an average SE count per tCNOT of 0.8 instead of the \added{ideal} 0.5\removed{we would aim for} in a deep circuit, and (2) the double-column logical layout, which was chosen to reduce shuttling distance, makes each SE round twice as slow.}

\removed{Reducing the spatial code distance in one direction as proposed in Ref. \cite{litinski_magic_2019} could allow a lower number of \snaq{} ancilla waves, further reducing its runtime. The specific code distances chosen would depend on the error rate of the noisy Ts, which could be close to the physical gate error rate if fast injection is used \cite{li_magic_2015}, or far lower if cultivation techniques are used first \cite{gidney_magic_2024}. Modeling cultivation on the \snaq{} architecture is beyond the scope of this work but is an important future step to reduce the cost of preparing high-quality T states.}

\subsubsection{8T-to-CCZ factory}

\added{The 8T-to-CCZ factory \cite{jones_lowoverhead_2013, gidney_efficient_2019} (Figure~\ref{fig:benchmarks}b) is an alternative magic state preparation method that outputs a $\ket{\text{CCZ}}$ state, which can be used to perform a Toffoli gate. We follow the transversal implementation from Ref. \cite{zhou_resource_2025} and compare to the lattice surgery implementation from Ref. \cite{litinski_magic_2019}, which involves 8 lattice surgery operations. For both transversal and lattice surgery implementations, we assume the input T states are prepared via cultivation \cite{gidney_magic_2024}, which is approximately 15$\times$ lower-cost on transversal-equipped architectures \cite{sahay_foldtransversal_2025} (accounting for the added shuttling overhead, which is small on \snaq{}). This results in a 4.3-5.6$\times$ more efficient $\ket{\text{CCZ}}$ factory at $d=27$, as shown in Table~\ref{tab:primitives}.}

\subsubsection{GHZ states}
\added{An $n$-qubit GHZ state (Figure~\ref{fig:benchmarks}d) can either be prepared using linear-nearest neighbour parallel tCNOTs \cite{yang_harnessing_2024} or implicitly created when performing a Pauli product measurement via lattice surgery, which we use as baseline. GHZ states are used in CNOT fan-outs, which, together with Toffoli gates (Figure~\ref{fig:benchmarks}c), form the Lookup subroutine, which is important for factoring \cite{gidney_how_2025} and chemistry simulation tasks that require PREP and SELECT oracles \cite{babbush_encoding_2018}. GHZ states are also used in Pauli exponentials in Trotterized Hamiltonian simulation \cite{lloyd_universal_1996}. \snaq{}'s GHZ state generation is 2.9-5.6$\times$ more efficient than the baselines.} 

\subsubsection{Logical qubit transport}
\added{The final primitive we study is the transport of logical information across large distances. A long-range transport operation can either be done by shuttling with intermediate SE rounds to avoid the accumulation of errors, or through lattice surgery \cite{litinski_game_2019}. Therefore, the latency of a transport operation increases with distance, but is upper-bounded by the cost of an LS operation. Assuming we perform one SE round for every 100 qubit distances traveled (Fig. \ref{fig:tcnot_vs_ls}), $k$-patch transport via shuttling in \snaq{} takes $\lfloor kd/100\rfloor \cdot t_{\text{SE}} + 2kd \cdot t_\text{sh}$ time. The SE round cost has been amortized for the result shown in Table~\ref{tab:primitives}.}

\subsection{Subroutines}\label{sec:subroutines}

\added{The two subroutines we consider are built of basic logical operations and the primitives discussed previously. We simulate the execution of their circuits, accounting for the previously calculated costs of primitives.}

\subsubsection{Adder}
\added{The quantum adder (Figure~\ref{fig:benchmarks}e) performs a reversible addition. It can be implemented as repeated applications of the MAJ and UMA blocks, which require $\ket{\text{CCZ}}$ states and local CNOT interactions. We use a transversal adder for SNAQ \cite{zhou_resource_2025}, and a volume-optimized adder using lattice surgery as the baseline \cite{gidney_how_2025}. The duration of both adders depends on the \textit{reaction time}, or how quickly the architecture can resolve a series of measurements where each measurement outcome determines the following measurement basis. Due to the need for a reactive CZ gate and measurement serialization, the SNAQ adder incurs a $7\times$ longer reaction time than the baseline, leading to a $2\times$ slower overall duration. However, due to cheaper $\ket{\text{CCZ}}$ factories and lower $\ket{\text{CCZ}}$ consumption rates, SNAQ achieves a lower volume with a much smaller $\ket{\text{CCZ}}$ factory, leading to a $3.2\times$ reduction in cost.}

\subsubsection{Lookup}
\added{The quantum Lookup (Figure~\ref{fig:benchmarks}f) appears in many algorithms due to its use in both quantum arithmetic \cite{gidney_windowed_2019} and chemistry \cite{babbush_encoding_2018}. It consists of a Toffoli ladder, implemented using $\ket{\text{CCZ}}$ states, and a CNOT-fanout that can be implemented using either a $\ket{GHZ}$ state or lattice surgery. The state-of-the-art transversal implementation is given in \cite{zhou_resource_2025}, and we use an established lattice surgery construction for the baselines \cite{gidney_how_2025, gidney_flexible_2019}. While the Toffoli part of the Lookup is reaction-time limited, this cost is hidden by the much costlier CNOT-fanout operation, which can be done in parallel; as such, the look-up is said to be Clifford-limited \cite{gidney_flexible_2019}. Due to faster transversal implementation of the CNOT-fanout, SNAQ achieves a $4.87\times$ to $6.26\times$ reduction in duration compared to baseline, leading to a $4.37\times$ to $5.74\times$ volume reduction.}

\added{\subsection{Algorithm impact}}

\added{We can heuristically estimate the impact on large-scale algorithms by considering the fraction of total algorithm volume spent on the above subroutines and the speedup achieved for that subroutine. The key subroutine for factoring is the lookup-controlled adder (CADD), which performs a quantum addition on states in specific superpositions prepared using a Lookup (Figure~\ref{fig:benchmarks}g). With one additional round of qubit shuttling, the data qubits can be laid out as illustrated in Figure~\ref{fig:benchmarks}g. The Lookup and Adder sizes are chosen to jointly optimize compilation cost and logical error-rate performance under hardware constraints such as reaction time and gate error rates. Although sweeping this full parameter space to determine the optimum for various factoring problems is an important direction, it is beyond the scope of this work. Instead, we use the optimized hyperparameters reported in \cite{gidney_how_2025, zhou_resource_2025} as reasonable reference points to estimate the range of improvements that SNAQ may achieve relative to the baselines, as shown in Table~\ref{tab:merged-subroutines}. 
Aside from two QFTs that contribute negligible volume, the entire factoring algorithm can be seen as repeated applications of CADD, which therefore account for the overwhelming majority of the total volume \cite{gidney_how_2025, zhou_resource_2025}. We therefore estimate that the 2.43-4.22$\times$ improvements in CADD (Table~\ref{tab:merged-subroutines}) propagate to very similar volume reductions for the entire factoring algorithm relative to baselines. This estimate accounts for all required qubit transport costs, which are included in the CADD cost.}

\added{Algorithms for chemistry simulation usually employ either a qubitization \cite{low_hamiltonian_2019, harrigan_expressing_2024} or Trotterization \cite{lloyd_universal_1996} approach. The state-of-the-art qubitization compilation technique uses repeated applications of the PREPARE and SELECT oracle \cite{babbush_encoding_2018}, where Lookup operations and single qubit rotations that require $\ket{T}$ states account for $\geq 90\%$ the cost\cite{lee_even_2021, zhou_resource_2025}. On the other hand, Trotterization employs repeated applications of Pauli exponentials, which are implementable using GHZ states and magic states.
We could expect time and volume reductions for both approaches, since \snaq{} provides speedups to $\ket T$ distillation, GHZ state preparation, and Lookups. However, the costs we estimate in Table~\ref{tab:merged-subroutines} are specific to the register sizes needed for factoring; the exact algorithmic speedup will depend on the specific chemistry problem being studied.}

\section{Conclusion}\label{sec:conclusion}

In this work, we proposed the \snaq{} surface code, a novel surface code implementation tailored for spin qubits. To overcome the readout component size problem, we introduced the idea of serialized ancilla qubit initialization and readout, showing that this is an effective way to enable error correction on a near-term-manufacturable silicon quantum dot array. \snaq{} is more space-efficient than prior proposals while providing a substantial logical clock speedup through transversal logic by leveraging rapid spin shuttling and a dense dot array. 


\added{\snaq{} is sensitive to coherence time but tolerates higher shuttling errors than the baselines, demonstrating the tradeoff between readout efficiency and qubit density.} \added{Concatenating \snaq{} with a higher-level QEC code could provide a path towards achieving ultra-low logical error rates beyond the coherence-limited error floor. The fixed-width spin qubit array may also be amenable to other specific classes QEC codes, in particular Floquet codes \cite{hastings_dynamically_2021} or adaptive concatenated schemes \cite{berthusen_adaptive_2025}. These codes may be harmed less by the serialization of ancilla measurement due to their already-partitioned syndrome extraction schedules.}

Overall, \snaq{} demonstrates that the often-assumed 1:1 readout-to-qubit ratio is neither necessary nor optimal on a shuttling-equipped architecture such as silicon spin qubits. Enabled by the unique capability of fast spin shuttling, \snaq{} is both orders-of-magnitude more area-efficient and significantly faster \added{for common subroutines} than proposals that mimic 2D grid layouts. Our work highlights qubit coherence and readout density as the most critical device metrics to enable this new architecture, providing a compelling, practical path toward fault-tolerant spin qubit processors.

\section*{Acknowledgements}

The authors thank Prithvi Prabhu, Fahd Mohiyaddin, and Nathaniel Bishop at Intel for helpful discussions on spin qubit manufacturing constraints and performance metrics. The authors are also grateful to Mariesa Teo for her assistance in understanding the significance of the adder and QROM benchmarks.

This work is funded in part by the STAQ project under award NSF Phy-232580; in part by the US Department of Energy Office of Advanced Scientific Computing Research, Accelerated 
Research for Quantum Computing Program; and in part by the NSF Quantum Leap Challenge Institute for Hybrid Quantum Architectures and Networks (NSF Award 2016136), in part by the NSF National Virtual Quantum Laboratory program, in part based upon work supported by the U.S. Department of Energy, Office of Science, National Quantum 
Information Science Research Centers, and in part by the Army Research Office under Grant Number W911NF-23-1-0077. The views and conclusions contained in this document are those of the authors and should not be interpreted as representing the official policies, either expressed or implied, of the U.S. Government. The U.S. Government is authorized to reproduce and distribute reprints for Government purposes notwithstanding any copyright notation herein.

FTC is the Chief Scientist for Quantum Software at Infleqtion and an advisor to Quantum Circuits, Inc.

\bibliographystyle{unsrt}
\bibliography{references}

\end{document}